\documentclass{nature}

\usepackage[numbers,super]{natbib}
\usepackage{graphicx}
\usepackage{booktabs}
\usepackage[usenames,dvipsnames,svgnames,table]{xcolor}
\usepackage{array}
\usepackage{booktabs}
\usepackage{hyperref}
\usepackage{amsmath,amssymb,amsfonts}
\usepackage{bm}
\usepackage{multicol}
\usepackage{multirow}
\usepackage{threeparttable}
\usepackage{url}
\usepackage{textcomp}
\usepackage{lineno}
\usepackage{booktabs}
\usepackage{csquotes}
\usepackage{multirow}
\usepackage{mathtools}
\usepackage{makecell}
\usepackage[normalem]{ulem}
\usepackage[font=footnotesize]{caption}
\usepackage{subcaption}
\usepackage{xspace}
\usepackage{amsthm}
\usepackage{dsfont}
\usepackage{xcolor}
\usepackage[normalem]{ulem}
\usepackage{float}
\usepackage{flafter} 

\makeatletter
\DeclareRobustCommand\onedot{\futurelet\@let@token\@onedot}
\def\@onedot{\ifx\@let@token.\else.\null\fi\xspace}
\def\eg{\emph{e.g}\onedot} 

\def\ie{\emph{i.e}\onedot}

\def\etal{\emph{et al}\onedot}
\makeatother
\DeclareMathOperator{\Loss}{\mathcal{L}}

\renewcommand{\vec}[1]{{\mathbf #1}}

\usepackage{amsthm}
\usepackage{pifont}
\usepackage{hyperref}
\hypersetup{
    colorlinks=true,
    linkcolor=blue,
    filecolor=magenta,      
    urlcolor=cyan,
}
\usepackage{lineno}

\usepackage[ruled,vlined]{algorithm2e}
\SetKwInput{KwInput}{Input}               
\SetKwInput{KwOutput}{Output} 

\usepackage[table]{xcolor}
\definecolor{headerblue}{RGB}{230,240,255}
\definecolor{rowgray}{gray}{0.96}

\usepackage{colortbl}  
\usepackage{tcolorbox}
\usepackage[capitalize]{cleveref}
\crefname{section}{Sec.}{Secs.}
\Crefname{section}{Section}{Sections}
\Crefname{table}{Table}{Tables}

\def\vtheta{{\bm{\theta}}}

\title{Segment Any Tumour: An Uncertainty-Aware Vision Foundation Model for Whole-Body Analysis}

\author{
Himashi Peiris$^{1}$,
Sizhe Wang$^{1}$,
Gary Egan$^{2}$,
Mehrtash Harandi$^{3}$,
Meng Law$^{4,5}$,
Zhaolin Chen$^{1,2,*}$
}

\begin{document}

\maketitle

\begin{affiliations}
\item Department of Data Science \& Artificial Intelligence, Faculty of Information Technology, Monash University, Melbourne, Australia.
\item Monash Biomedical Imaging (MBI), Monash University, Melbourne, Australia.
\item Department of Electrical \& Computer Systems Engineering, Faculty of Engineering, Monash University, Melbourne, Australia.
\item Department of Neuroscience, The School of Translational Medicine, Monash University, Melbourne, Australia.
\item Department of Radiology, Alfred Health, Melbourne, Australia.
\end{affiliations}

\noindent\text{${^*}$Corresponding Author(s) E-mail(s): Zhaolin.Chen@monash.edu}

\section*{Abstract}
\label{abstract}

\begin{abstract}
Prompt-driven vision foundation models, such as the Segment Anything Model, have recently demonstrated remarkable adaptability in computer vision. However, their direct application to medical imaging remains challenging due to heterogeneous tissue structures, imaging artefacts, and low-contrast boundaries, particularly in tumours and cancer primaries leading to suboptimal segmentation in ambiguous or overlapping lesion regions.
Here, we present Segment Any Tumour 3D (SAT3D), a lightweight volumetric foundation model designed to enable robust and generalisable tumour segmentation across diverse medical imaging modalities. SAT3D integrates a shifted-window vision transformer for hierarchical volumetric representation with an uncertainty-aware training pipeline that explicitly incorporates uncertainty estimates as prompts to guide reliable boundary prediction in low-contrast regions. Adversarial learning further enhances model performance for the ambiguous pathological regions.
We benchmark SAT3D against three recent vision foundation models and nnUNet across 11 publicly available datasets, encompassing 3,884 tumour and cancer cases for training and 694 cases for in-distribution evaluation. Trained on 17,075 3D volume-mask pairs across multiple modalities and cancer primaries, SAT3D demonstrates strong generalisation and robustness. To facilitate practical use and clinical translation, we developed a 3D Slicer plugin that enables interactive, prompt-driven segmentation and visualisation using the trained SAT3D model. Extensive experiments highlight its effectiveness in improving segmentation accuracy under challenging and out-of-distribution scenarios, underscoring its potential as a scalable foundation model for medical image analysis.
\end{abstract}

\begin{figure*}[h!]
    \centering
    \includegraphics[width=1.0\linewidth]{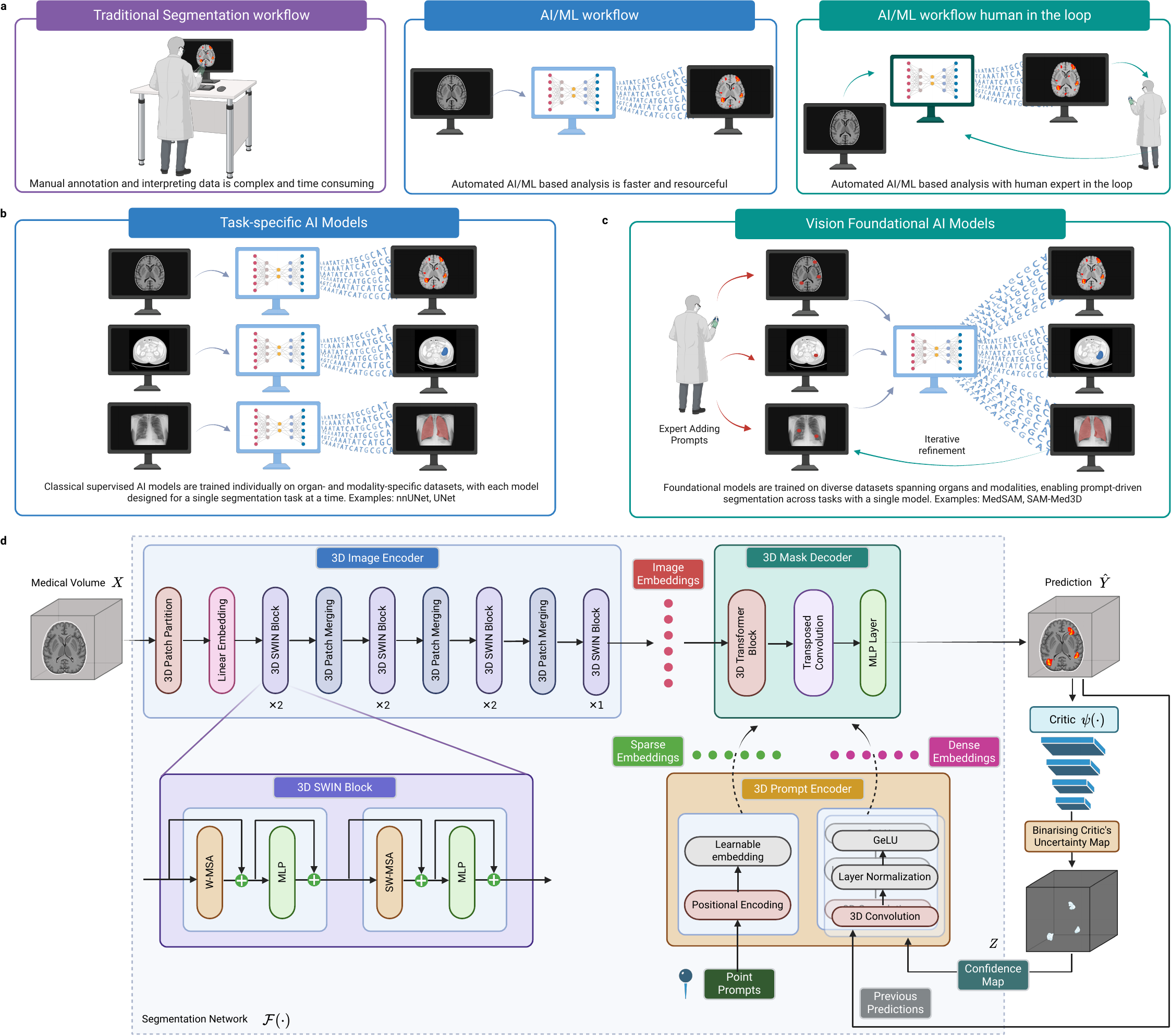}
    \caption{\scriptsize \textbf{Segmentation Workflows, Model Paradigms, and the Proposed Vision Foundation Model.} 
\textbf{a.} Illustration depicts three segmentation workflows: traditional manual segmentation, AI/ML-based automated segmentation, and AI/ML-based segmentation with human-in-the-loop refinement. 
\textbf{b.} Task-specific deep learning paradigms. Here, predictions are produced in a fully automated manner without provision for human intervention or adjustment, reflecting the deterministic and task-locked nature of such models.
\textbf{c.} Prompt-driven vision foundation models. Here, predictions are conditioned on user-provided prompts, allowing flexible adaptation across diverse tasks and modalities, and enabling human-in-the-loop interaction to refine segmentation outcomes.
\textbf{d.} Overview of the proposed uncertainty-aware vision foundation architecture. Here, $\mathcal{F}(\cdot)$ represents the SAT3D segmentation backbone, while $\psi(\cdot)$ denotes its discriminator (critic) network. For an input medical volume $X$, the prediction mask $\hat{Y}$ is generated by $\mathcal{F}(\cdot)$ and subsequently evaluated by $\psi(\cdot)$ to produce a confidence map $Z$, highlighting certain and uncertain regions.}
    \label{fig:figure_1}
\end{figure*}

Accurate segmentation of tumours and lesions is a cornerstone of clinical imaging workflows, supporting diagnosis, treatment planning, and longitudinal monitoring. The~\cref{fig:figure_1}a and  \cref{fig:figure_1}b provide an overview of the progression in segmentation workflows and model paradigms. Traditionally, this process has relied on manual annotation or interpretation by radiologists or medical professionals, a time-consuming, subjective task with substantial inter-observer variability. The emergence of Artificial Intelligence (AI), Machine Learning (ML), and Deep Learning (DL) models has accelerated segmentation workflows, particularly for organ-specific and modality-specific applications, such as liver tumours, kidney tumours, or lung cancer~\cite{msd}. These task-specific models, such as nnUNet~\cite{nnunet}, are typically trained on dedicated datasets and optimised for a single anatomical or pathological target~\cite{nnunet,ronneberger2015u,sinclair2024perivascular}. nnUNet, for example, has become the de facto standard baseline in medical image segmentation, automatically configuring preprocessing, network architectures, and training pipelines for a given dataset, consistently achieving state-of-the-art performance across various tasks. Consequently, these models perform well within training datasets; however, they often lack generalisability since they are trained under controlled conditions, \eg, specific organs or pathologies, and their performance deteriorates on unseen lesion types, rare presentations, or out-of-distribution datasets, where these challenges are common in real-world medical imaging.
Furthermore, fully automated task-specific models are often end-to-end optimised, and lack a clinician-in-the-loop design, where medical professionals can interactively refine or validate AI-generated segmentations. Clinician-in-the-loop can especially be valuable in scenarios involving ambiguous boundaries, low-contrast lesions, or unusual anatomical variations. These systems enhance trust, ensure safety, and improve outcomes by combining the efficiency of automation with the expertise of clinicians~\cite{campanella2025real,zhang2023segment}. 

Recent advances in Vision Foundation Models (VFMs), such as the Segment Anything Model (SAM), have introduced a prompt-driven paradigm in vision tasks~\cite{SAM,ravi2024sam}. These models are pre-trained on large-scale datasets and can generalise across object boundaries using human inputs such as points, bounding boxes, or masks. 
They also demonstrate strong generalisation across a wide range of natural image segmentation tasks, but their application to medical imaging often yields suboptimal results~\cite{MedSAM,sam_survey,ma2024segment,zhu2024medical}. This performance gap arises due to several factors: (i) the domain shift between natural and medical images, (ii) the subtle and ambiguous boundaries of many lesions, and (iii) the lack of medical-specific context in the pretraining datasets. 
To mitigate these limitations, several studies have improved and extended the SAM architecture, including MedSAM~\cite{MedSAM}, SAM-Med2D~\cite{SAMMED2D}, SAM-Med3D~\cite{SAMMED3D}, and SAM3D~\cite{SAM3D}. These methods adapt the SAM architecture by fine-tuning its image encoder or prompt encoder on medical images, incorporating 3D volumetric support, and leveraging point-based or box-based prompts to guide segmentation across imaging modalities.
Beyond prompt-driven methods~\cite{ye2023uniseg}, recent studies have advanced toward foundation models for medical imaging, pre-trained across large-scale, multi-institutional datasets to achieve generalisable representation learning. These efforts include generalist models such as RadFM~\cite{RADFM} and MedSegX~\cite{zhang2025generalist}, designed to capture broad anatomical and modality coverage, domain-specialised models such as MRI-PTPCa~\cite{MRIPTPCA} and GPFM~\cite{GPFM}, which focus on disease-specific interpretation, and task-adaptive models like META-SiM~\cite{METASIM}, which leverage multitask pretraining for efficient transfer learning. Among these, MedSegX introduced a large-scale generalist segmentation framework trained on 1.67 million image–mask pairs from 134 datasets (10 modalities, 39 organs) using a hierarchical ontology (MedSegDB) to enhance contextual representation learning.
In contrast, SAT3D focuses on volumetric tumour segmentation, trained from scratch on a specialised multimodal tumour dataset. It integrates uncertainty-aware prompt guidance with adversarial optimisation to improve reliability under low-contrast and ambiguous tumour boundaries.
Unlike MedSegX, which performs slice-wise 2D pretraining, SAT3D processes entire 3D scan volumes, capturing complete tumour morphology and inter-slice continuity to mitigate spatial discontinuities and loss of fine-structure detail commonly observed in stacked 2D predictions. Building on the success of the SAM and its recent medical adaptations, which remain state-of-the-art in prompt-driven segmentation, SAT3D extends these foundations to volumetric medical data. While SAM-based frameworks have demonstrated strong generalisation across diverse medical datasets, real-world clinical applications still pose challenges, particularly in modelling the complex and subtle appearance of tumours and cancer primaries. Variations in tumour size, shape, and texture introduce additional difficulty, especially in primary cancer detection or whole-body lesion segmentation~\cite{ali2024evaluating}.

To address these limitations, we re-examine and refine the SAM architecture for medical imaging, with a focus on the complex task of segmenting tumours and cancer primaries. We introduce a lightweight and robust volumetric VFM for tumour and cancer primary segmentation, trained on public data repositories termed SAT3D, which integrates uncertainty estimation into the architecture through both discriminative modelling with adversarial learning and prompt-based uncertainty guidance, enabling more robust predictions in regions of low contrast or ambiguous boundaries. 
We further curate a diverse, volumetric dataset which comprises 11 publicly available datasets (Automated Lesion Segmentation in Whole-Body PET/CT (AutoPET 2024)~\cite{gatidis2023autopet,ingrisch_2024_10990932}, Head and Neck Tumor Segmentation for MR-Guided Applications (HNTSMRG 2024)~\cite{wahid_2024_11199559,wahid2024overview}, Tumor Detection, Segmentation and Classification Challenge on Automated 3D Breast Ultrasound (TDSC-ABUS 2023)~\cite{luo2025tumor}, Kidney PArsing Challenge 2022 (KiPA 2022)~\cite{guanyu_yang_2022_6361938}, Kidney Tumor Segmentation Challenge 2023 (KiTS 2023)~\cite{heller2019kits19,heller2023kits21}, Liver Tumor Segmentation Challenge (LiTS)~\cite{BILIC2023102680}, Medical Segmentation Decathlon Challenge (MSDC) Lung~\cite{msd}, MSDC Colon~\cite{msd}, MSDC Pancreas~\cite{msd}, MSDC Hepatic Vessel~\cite{msd}, Brain Tumour Segmentation Challenge 2021 (BraTS 2021)~\cite{baid2021rsna,menze2014multimodal,bakas2017advancing,bakas2017segmentation,bakas2017segmentation1}) encompassing a wide spectrum of tumour types including: head and neck-associated Gross Tumour Volume of the primary Tumour (GTVp) and Gross Tumour Volume of Lymph Nodes (GTVn); brain tumour subregions such as Surrounding Non-Enhancing Fluid-Attenuated Inversion Recovery (FLAIR) Hyperintensity (SNFH) often call it as Brain swelling or Edema, enhancing tumour, and Necrotic and Non-Enhancing Tumour Core (NCR); upper-body cancers such as lung cancer nodules and breast tumours; abdominal and pelvic primaries including hepatic tumours, renal tumours, liver tumours, pancreatic cancers, kidney tumours, and colon cancers; as well as whole-body metabolically active tumour lesions and anatomical locations across the body such as Brain, Head, Neck, Upper Body, Abdomen and whole body covering medical imaging modalities such as Magnetic Resonance Imaging (MRI) (including sequences T1-weighted, T2-weighted, FLAIR, contrast-enhanced T1-weighted), Computed Tomography (CT), Computed Tomography Angiography (CTA), fluorodeoxyglucose (FDG) Positron Emission Tomography (PET) and Ultrasound, supporting generalisable training and evaluation.

We conduct a large-scale study to evaluate the performance of the proposed SAT3D model, benchmarking it against recent vision foundation models for medical imaging, including SAM-Med3D~\cite{SAMMED3D}, SAM-Med3D (Turbo)~\cite{SAMMED3D}, and FastSAM3D~\cite{shen2024fastsam3d}. SAT3D demonstrated significant improvements both quantitatively and qualitatively over these foundation models. We further compared SAT3D with nnUNet, the most widely adopted task-specific segmentation model in medical imaging, and found that SAT3D performs on par with nnUNet~\cite{nnunet}. In addition, experiments on out-of-distribution datasets demonstrate SAT3D’s strong generalisability. To facilitate practical translation, we developed a preliminary 3D Slicer plugin~\cite{pieper20043d} that enables interactive visualisation and prompt-based segmentation using SAT3D, laying the foundation for future integration into clinical workflows.

\begin{figure*}[ht!]
    \centering
    \includegraphics[width=0.93\linewidth]{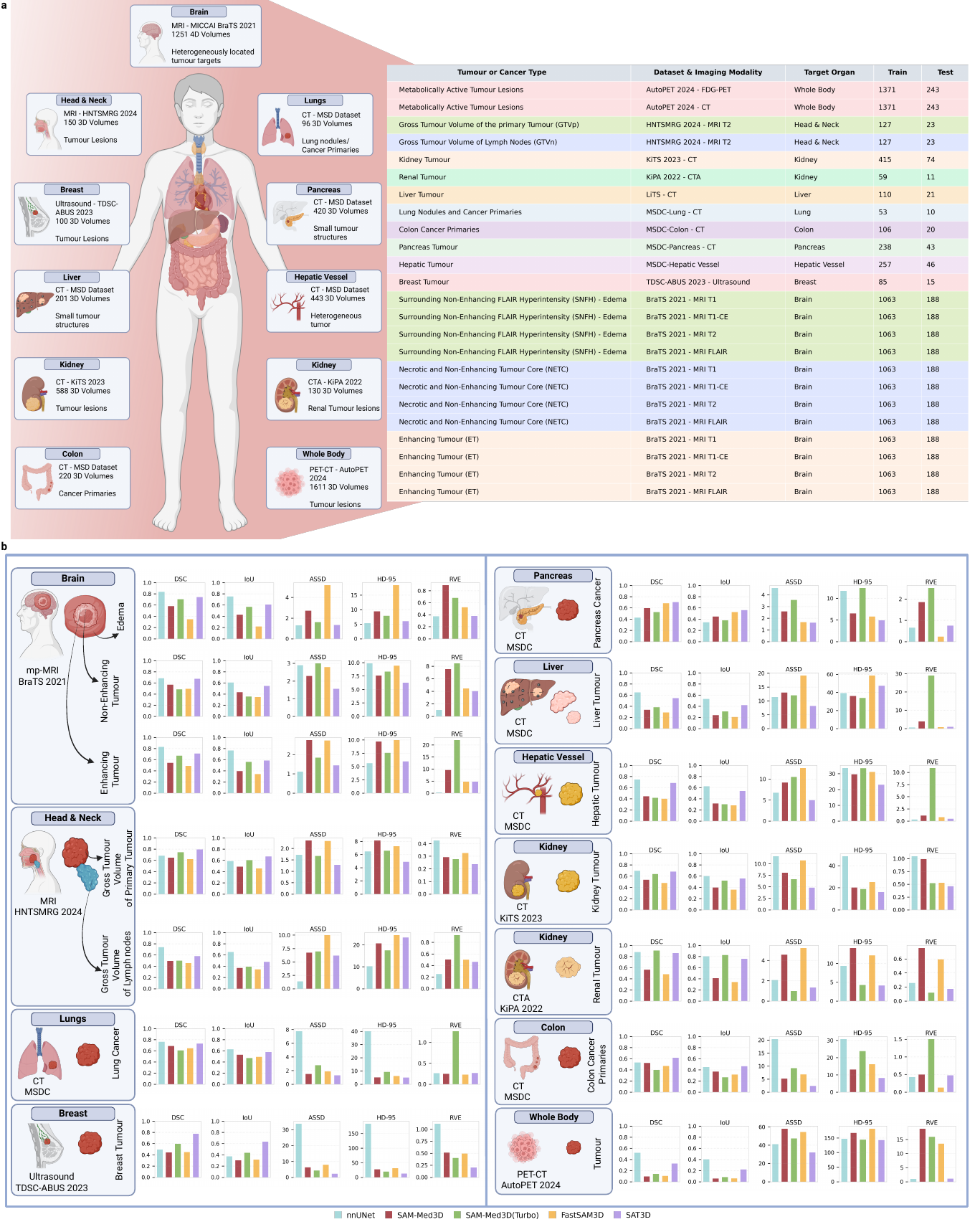}
    \caption{\scriptsize \textbf{Overview of the datasets and Performance Evaluation on In-Distribution data.} \textbf{a.} The SAT3D model is trained on a large-scale, diverse collection of medical imaging datasets encompassing various anatomical regions, pathological tumour types, and imaging modalities. \textbf{b.} The comparison includes the task-specific nnUNet, SAM-based models for 20-point prompts (SAM-Med3D, SAM-Med3D(Turbo) and FastSAM3D) and ours, SAT3D. Performance is reported in bar plots across five metrics: DSC, IoU, ASSD, HD-95, and RVE.}
    \label{fig:figure_2}
\end{figure*}

\section*{Results}
The goal of the SAT3D model was to establish a more robust vision foundation model capable of accurately detecting and segmenting critical pathological regions, such as tumours and cancer primaries. A key challenge in designing such models lied in achieving high performance across a wide variety of tumour types and anatomical sites, while effectively handling the diversity of medical imaging modalities, acquisition protocols, artefacts, and complex pathological variations. 

In this study, we revisited and extended recent architectural improvements, specifically, models like SAM-Med3D, to enhance generalisability in more demanding segmentation tasks involving tumours/cancer primaries. We trained and evaluated our method across a broad spectrum of tumour and cancer types, including: head and neck-associated GTVp and GTVn regions; brain tumour subregions such as edema, enhancing tumour, and non-enhancing tumour (necrosis); upper-body cancers such as lung and breast cancer; abdominal and pelvic primaries, including hepatic, renal, liver, pancreatic, and colon cancers; as well as whole-body metabolically active tumour lesions. To support learning across these diverse clinical scenarios, our dataset spanned multiple imaging modalities, including T1-weighted, T2-weighted, FLAIR, and T1 contrast-enhanced MRI, as well as FDG-PET, CT, CTA, and ultrasound.
For evaluation, we compared the performance of the proposed SAT3D model with several recent vision foundation models, including SAM-Med3D~\cite{SAMMED3D}, SAM-Med3D (Turbo)~\cite{SAMMED3D}, and FastSAM3D~\cite{shen2024fastsam3d}.
SAM-Med3D was pretrained on approximately 150,000 3D medical volumes collected from over 80 public datasets, covering diverse modalities such as CT, MRI, PET, and ultrasound.
The SAM-Med3D (Turbo) variant further expanded its pretraining to nearly 500,000 volumetric scans across more than 120 datasets.
In contrast, FastSAM3D employed a more compact dataset of around 45,000 scans, prioritising inference efficiency through architectural simplification.
For comparison with a task-specific baseline, we trained nnUNet~\cite{nnunet} from scratch using the same custom splits employed for SAT3D, ensuring identical data partitions and experimental fairness.
Our proposed SAT3D was trained on 17,075 3D volume–mask pairs spanning 11 publicly available datasets across PET, CT, MRI, CTA, and ultrasound modalities, encompassing 3,884 tumour and cancer cases for training and 694 cases for in-distribution evaluation.

Unlike nnUNet, which is a task-specific model trained from scratch on a single dataset to optimise performance for a defined segmentation task, foundation models such as SAT3D and SAM-Med3D are pretrained across large, heterogeneous datasets spanning multiple organs, modalities, and tumour types.
This large-scale, multi-domain pretraining enables the learning of generalisable representations that can be transferred to unseen tasks through prompting or zero-shot inference, without requiring task-specific retraining.
In contrast, nnUNet’s data-specific optimisation provides strong in-distribution performance but limited generalisability beyond its training cohort~\cite{ma2023towards}.
Unlike conventional vision foundation models that depend on massive, web-scale datasets, SAT3D achieves foundation-level generalisation through efficient architectural design and uncertainty-guided prompt learning rather than sheer data volume. This design philosophy aligns with recent insights from dataset distillation and data-efficient learning~\cite{lu2025general}, suggesting that carefully curated and informative samples can yield comparable representational quality to large, redundant datasets.
Thus, SAT3D’s advantage lies in its ability to leverage foundation-level representations while maintaining task adaptability through uncertainty-guided prompt learning, bridging the gap between generalist foundation models and specialised clinical segmentation networks.

\paragraph{Analysis of the Segmentation Results on In-Distribution Data.}

\begin{figure*}[ht!]
    \centering
    \footnotesize
    \includegraphics[width=1.0\linewidth]{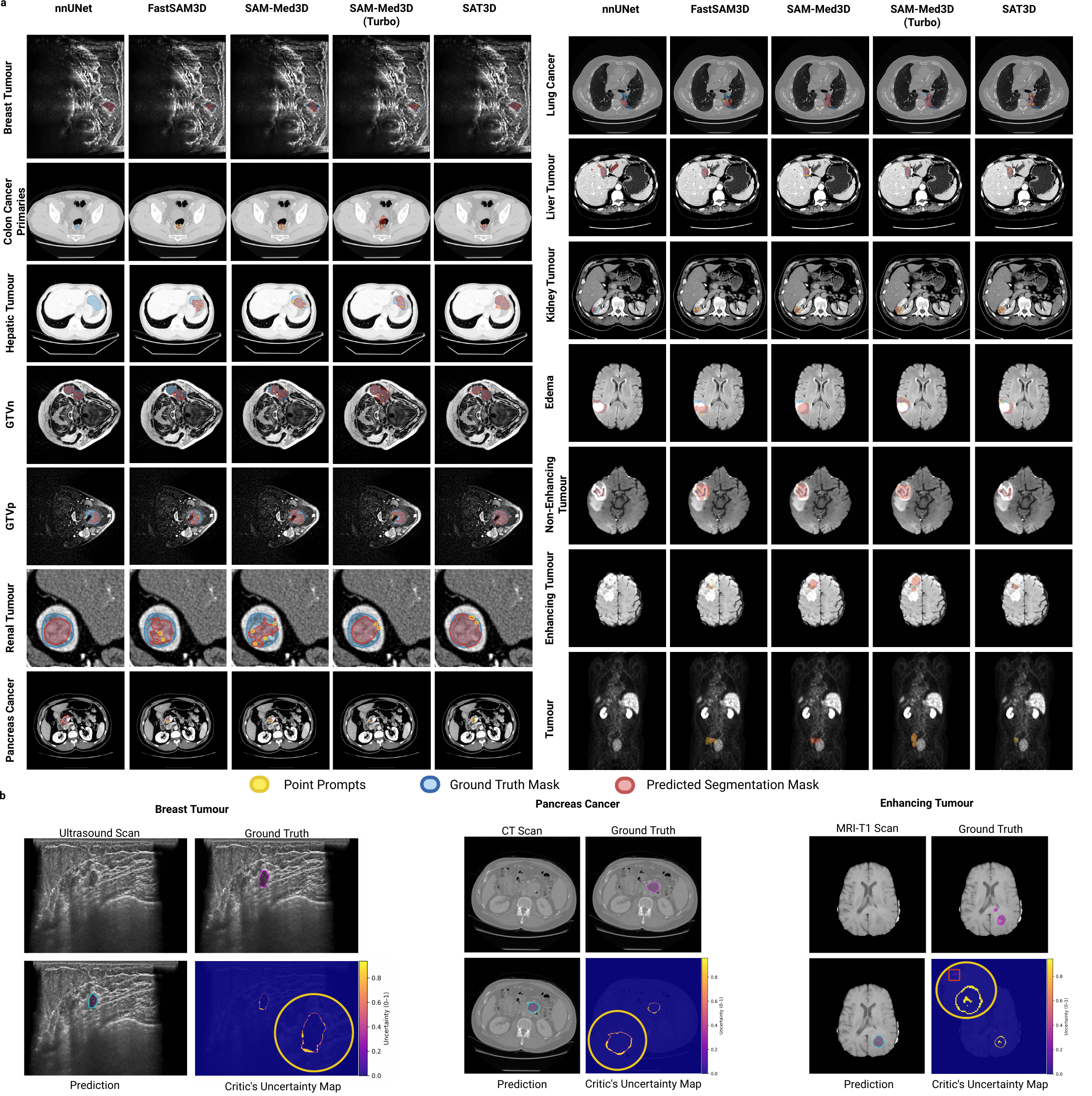}
    \caption{\scriptsize \textbf{Qualitative performance evaluation across tumours and cancer primaries.}  \textbf{a.} Comparison of Segmentation masks predicted from each method, including the task-specific nnUNet, SAM-based models for 20-point prompts (SAM-Med3D, SAM-Med3D(Turbo) and FastSAM3D) and ours, SAT3D (See Extended Data Fig. 1 for Qualitative Comparison for Volumetric Segmentations and Extended Data Fig. 2 for Qualitative performance Comparison on multi-modal imaging data). \textbf{b.} Uncertainty information captured by the critic network during the training of the SAT3D model. Here, yellow-coloured pixels indicate high-uncertainty regions, whereas blue pixels indicate low-uncertainty regions. Yellow outlined circles show a zoomed-in view of uncertain regions.}
    \label{fig:figure_3}
\end{figure*}

As shown in~\cref{fig:figure_2}b and \cref{fig:figure_3}a, when compared with other VFMs for medical imaging, SAT3D consistently outperformed all alternatives across evaluation metrics. For the Dice Similarity Coefficient (DSC), SAT3D achieved a mean of 0.672, which is markedly higher than those of SAM-Med3D (0.503), FastSAM3D (0.457), and SAM-Med3D Turbo (0.550), confirming its superiority in volumetric accuracy. The advantage of SAT3D also extended to Intersection over Union (IoU), where it maintained higher overlap scores across tumour and organ boundaries, reflecting more reliable and stable delineation. 
Boundary-sensitive metrics further underscore SAT3D’s robustness. While SAM-Med3D and FastSAM3D suffered from significant errors in the Hausdorff Distance 95th Percentile (HD-95) and Average Symmetric Surface Distance (ASSD), highlighting unstable boundary predictions, SAT3D achieved substantially lower boundary distances, particularly for challenging tumours such as pancreas, renal, and head-and-neck lesions. For example, SAT3D reduced HD-95 by 5-10 mm compared to VFMs and consistently halved ASSD errors on small or irregular structures. Similarly, Relative Volume Error (RVE) analysis demonstrated that SAT3D mitigated volumetric bias by 20-40\% compared to VFMs, reducing the tendency of foundation models to over-segment ambiguous regions or underfit subtle lesions.

When directly compared with the task-specific nnUNet, SAT3D achieved a nearly identical mean Dice (0.672 vs. 0.676). Crucially, SAT3D matched or exceeded nnUNet on several of the most challenging datasets, including pancreas cancer, renal tumours, and head-and-neck (gtvp/gtvn), where it not only improved Dice but also achieved lower HD-95 and more stable RVE. These results emphasise SAT3D’s reliability in cases where precise tumour delineation is clinically most demanding. Nevertheless, nnUNet retained marginal advantages on large, high-contrast structures such as liver and lung, where it achieved slightly higher Dice and IoU, consistent with its strong volumetric optimisation.
Unlike SAT3D, nnUNet operates purely in a supervised setting without any prompt-based conditioning or interactive refinement. These results highlight SAT3D’s ability to leverage point-based and uncertainty-driven prompts as an additional advantage, enabling targeted refinement in ambiguous regions where conventional fully supervised models typically struggle.

It is important to note that for multi-modal datasets such as AutoPET 2024 (CT and FDG-PET) and BraTS 2021 (T1-w, T2-w, contrast-enhanced T1-w, and FLAIR), the task-specific nnUNet was trained jointly on all available modalities, treating them as multi-channel inputs during training and thereby exploiting their complementary information. In contrast, VFMs, including SAT3D, were trained with a single-modality data loader, without direct multi-modal integration. To ensure fairness in our primary experiments, we therefore reported for VFMs the best mean Dice score across modalities when benchmarking against nnUNet. In our extended analysis, however, we also present a modality-wise breakdown of performance to highlight how each modality contributes to the cancer and tumour segmentation task.

\cref{fig:figure_3}a, presented qualitative comparisons across multiple tumour and cancer types. The visualisation highlighted the predicted segmentation masks obtained from SAT3D alongside those from comparison methods such as nnUNet, FastSAM3D, SAM-Med3D and SAM-Med3D (Turbo). It can be seen that the SAT3D consistently demonstrated improved boundary delineation and better preservation of tumour morphography across the majority of cases, and on par performance with the task-specific model nnUNet in some cases. In contrast, baseline VFM methods often exhibited over-segmentation, under-segmentation or irregular boundaries. In~\cref{fig:figure_3}b, we also illustrated how the critic network's addition facilitates uncertainty-aware training of the SAT3D model. These results complement our quantitative findings (see~\cref{tab:table_1}).

\begin{figure*}[ht!]
    \centering
    \includegraphics[width=0.98\linewidth]{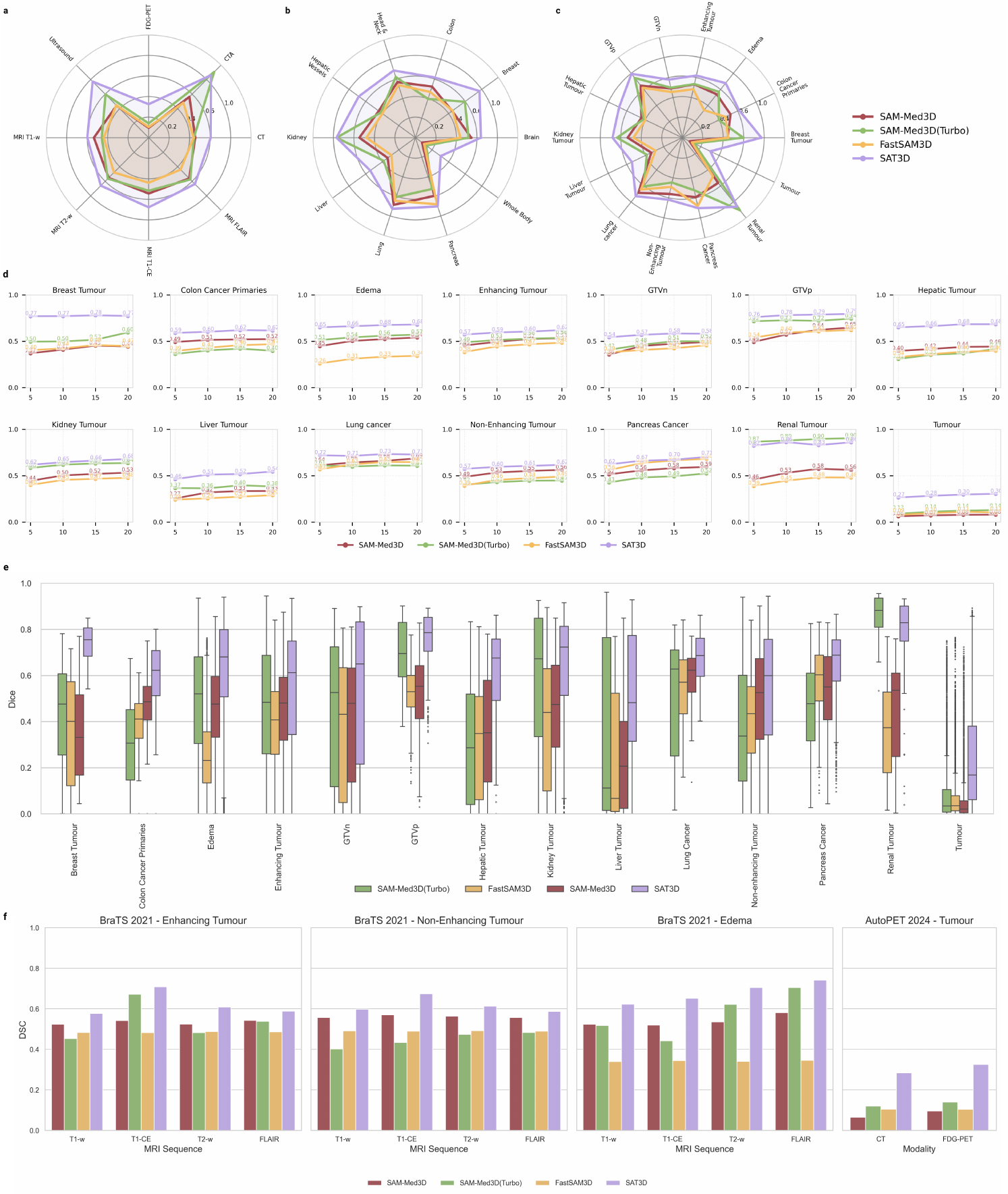}
    \caption{\scriptsize \textbf{Performance analysis of SAM-based models.} \textbf{a.} Imaging Modality-wise DSC score performance showing differential model behaviour across CT, MRI Sequences, CTA, FDG-PET and Ultrasound imaging for four SAM-based segmentation models. \textbf{b.} Organ-wise DSC score comparison. \textbf{c.} DSC score comparison across tumour/cancer types. \textbf{d.} DSC score trends as the number of point prompts increases across four SAM-based segmentation models. \textbf{e.} Distribution of DSC scores in box-and-whisker plots. \textbf{f.} Performance Analysis on Multi-modal datasets. 
    }
    \label{fig:figure_4}
\end{figure*}

Radar plots in~\cref{fig:figure_4}a, \cref{fig:figure_4}b and \cref{fig:figure_4}c, provided a holistic view of segmentation performance across imaging modalities, organs, and tumour/cancer primaries under the 20-point prompt setting. The modality-wise analysis showed that SAT3D consistently achieved superior Dice scores across CT, CTA, FDG-PET, ultrasound, and all MRI sequences, demonstrating strong adaptability to heterogeneous imaging sources. Organ-wise comparisons further underscored SAT3D’s consistent performance, particularly in abdominal and thoracic organs, where FastSAM3D and SAM-Med3D showed larger drops in accuracy. Tumour-wise analysis reinforced these findings: SAT3D achieved the highest Dice medians in the majority of cancer primaries, including breast, colon, hepatic, and renal tumours, and matched or outperformed SAM-Med3D(Turbo) in head-and-neck regions (GTVp, GTVn, edema). Overall, the radar plots emphasise that while all SAM-based foundation models benefit from prompts, SAT3D consistently delivers balanced and superior performance across diverse anatomical and modality domains.

We further analyzed how Dice scores fluctuated with the number of point prompts (5, 10, 15, and 20) across 14 tumour categories. The results in~\cref{fig:figure_4}d, revealed that all prompt-driven foundation models exhibited a consistent upward trend in performance with an increasing number of point prompts, though the magnitude of improvement varied across tumour/cancer types. SAT3D demonstrated the most stable and consistent gains, particularly in anatomically challenging cases such as liver, pancreas, and renal tumours. In contrast, FastSAM3D showed the largest fluctuations, with occasional drops in performance between prompt settings, suggesting limited robustness and higher sensitivity to prompt configuration. SAM-Med3D and SAM-Med3D(Turbo) generally followed smoother trajectories, though Turbo occasionally outperformed the base variant in large solid tumours. These findings highlight the role of prompts in enhancing segmentation accuracy, with SAT3D offering both the highest Dice scores and the least variability across tumour types.

Taken together, these findings show that although enriched prompting improves baseline VFMs, they remain hindered by unstable boundary metrics and volumetric inconsistencies. SAT3D bridges this gap, combining the cross-domain adaptability of foundation models with task-specific discriminative power. In doing so, it delivers performance competitive with nnUNet on average while decisively surpassing all other VFMs across DSC, IoU, HD-95, ASSD, and RVE.

\paragraph{Statistical Analysis \& Insights.}
\label{sec:stat}
A Friedman rank-sum test, as in~\cref{tab:table_1}, was conducted to assess whether the observed performance differences among the segmentation methods were statistically significant across tumour types and evaluation metrics. The test yielded a chi-square statistic of $\chi^2 = 89.54$ with an associated probability of $p < 1 \times 10^{-18}$, indicating an extremely low likelihood that the observed rank differences occurred by chance. This result provides strong evidence against the null hypothesis that all methods perform equivalently, confirming that the performance variations across methods are statistically meaningful.
From the \cref{tab:table_1}, it can be seen that the SAT3D consistently achieved the best performance with the lowest average rank (1.73), followed by nnUNet (2.46). In contrast, SAM-Med3D variants and FastSAM3D obtained higher ranks, indicating inferior performance. These results highlighted SAT3D’s superior robustness across tumour types and metrics, with nnUNet remaining competitive, while the SAM-based approaches lagged behind.
As shown in~\cref{fig:figure_4}e, the results also confirmed that prompt-driven foundation models (SAT3D and SAM-Med3D variants) significantly outperformed lighter baselines such as FastSAM3D in tumour segmentation accuracy, with SAT3D showing the most consistent gains across diverse tumour types.
We additionally conducted a non-parametric pairwise analysis of DSC distributions across all tumour types (see Extended Data Table 1), comparing SAT3D against nnUNet, SAM-Med3D, SAM-Med3D(Turbo), and FastSAM3D under the 20-point prompt configuration.
Wilcoxon signed-rank tests confirmed statistically significant differences ($p<0.05$) for the majority of tumour categories, indicating meaningful distributional shifts between methods.
Overall, SAT3D consistently achieved significantly higher DSC values than FastSAM3D and SAM-Med3D across most solid tumours, including hepatic, pancreatic, and renal lesions, while showing comparable or slightly lower performance than nnUNet in large, high-contrast organs such as liver and lung.
SAM-Med3D(Turbo) demonstrated competitive performance with SAT3D, particularly in brain-related regions (GTVp, GTVn, and edema), suggesting that both prompt-conditioned models adapt well to complex or low-contrast contexts.
The global Wilcoxon analysis across all tumour types yielded extremely low p-values ($1.05\times10^{-32}$, $2.99\times10^{-149}$, and $4.57\times10^{-152}$ for SAT3D vs. nnUNet, SAM-Med3D, and FastSAM3D, respectively), confirming significant methodological differences in segmentation behaviour across diverse anatomical contexts.

\paragraph{Performance Analysis on Multi-modal datasets.}
The combined analysis in~\cref{fig:figure_4}f highlighted consistent trends across both BraTS and AutoPET whole-body datasets. For brain tumours, SAT3D achieves the highest DSC scores across all MRI sequences for enhancing, non-enhancing, and edema regions, demonstrating robust performance relative to SAM-Med3D, SAM-Med3D(Turbo), and FastSAM3D. Notably, the performance gains were most pronounced for edema and non-enhancing regions, where boundary delineation was typically more challenging. In contrast, the whole-body tumour analysis revealed the same trend: SAT3D outperformed competing methods on both CT and FDG-PET modalities, while FastSAM3D showed comparatively weaker results. These findings underscore SAT3D’s ability to generalise across different tumour subtypes and imaging modalities.
However, absolute DSC scores for tumour detection in both CT and FDG-PET in the AutoPET 2024 dataset remain low. This limitation can be attributed to the restricted training diversity, as only one PET-related dataset and one whole-body dataset were available. When compared to nnUNet, the widely adopted task-specific segmentation model, there remained a significant performance gap for VFMs, including SAT3D. This underscores the current challenge of extending foundation model capabilities from well-represented neuro-oncology tasks to more diverse whole-body tumour detection settings, where richer and more balanced multimodal training data will be essential to close the gap.

\paragraph{Incorporating Segmentation Uncertainty as a Dense Prompt.} 
Randomness stemming from prompts could lead to variability in segmentation performance and algorithmic reliability. To mitigate this issue, in our study, we utilise a critic/discriminator network for voxel-wise true/false classification, and derive uncertainty information of the predicted segmentation mask. This uncertainty information is then used to generate a confidence map (See~\cref{fig:figure_1}d) and used as an 
additional dense prompt. Instead of relying solely on the previous mask generated for earlier point prompts as done in recent VFMs, we incorporate both the previous segmentation mask and its corresponding confidence map as dense prompts during training, enabling the model to refine predictions more reliably across successive interactions.
Uncertainty information captured by the critic network is visualised in~\cref{fig:figure_4}b. As illustrated, the regions of high uncertainty predominantly concentrate around the boundaries of the predicted segmentation masks. This behaviour is expected in medical imaging tasks, particularly for tumour and cancer delineation, where lesion borders are often irregular, heterogeneous, and poorly defined due to imaging artefacts, partial volume effects, or low contrast between pathological and healthy tissues.
By guiding the network to focus more on boundary regions with higher uncertainty, the framework improves its ability to capture small and detailed tumour regions. As shown in~\cref{fig:figure_3}e, the critic’s uncertainty map highlights additional areas of interest around the enhancing tumour in the brain MRI (marked by the red box), complementing the main segmentation result. These boundary-aware cues are especially useful in clinical settings, where identifying even small or subtle tumour regions can support more accurate staging, treatment planning, and prognosis.
Incorporating this uncertainty information, along with the prompts, into training provides valuable cues for the model. By leveraging uncertainty-aware learning, the network not only improves its boundary precision but also better accounts for inter-observer variability, which is common in clinical annotation of tumours across modalities like MRI, CT, and PET. Radiologists often disagree on the precise extent of tumour boundaries, especially in infiltrative cancers such as gliomas or head and neck primaries. The integration of uncertainty not only regularises the model but also improves trustworthiness by highlighting regions where predictions are less reliable, potentially guiding clinicians to review critical areas.

\begin{figure*}[ht!]
    \centering
    \includegraphics[width=1\linewidth]{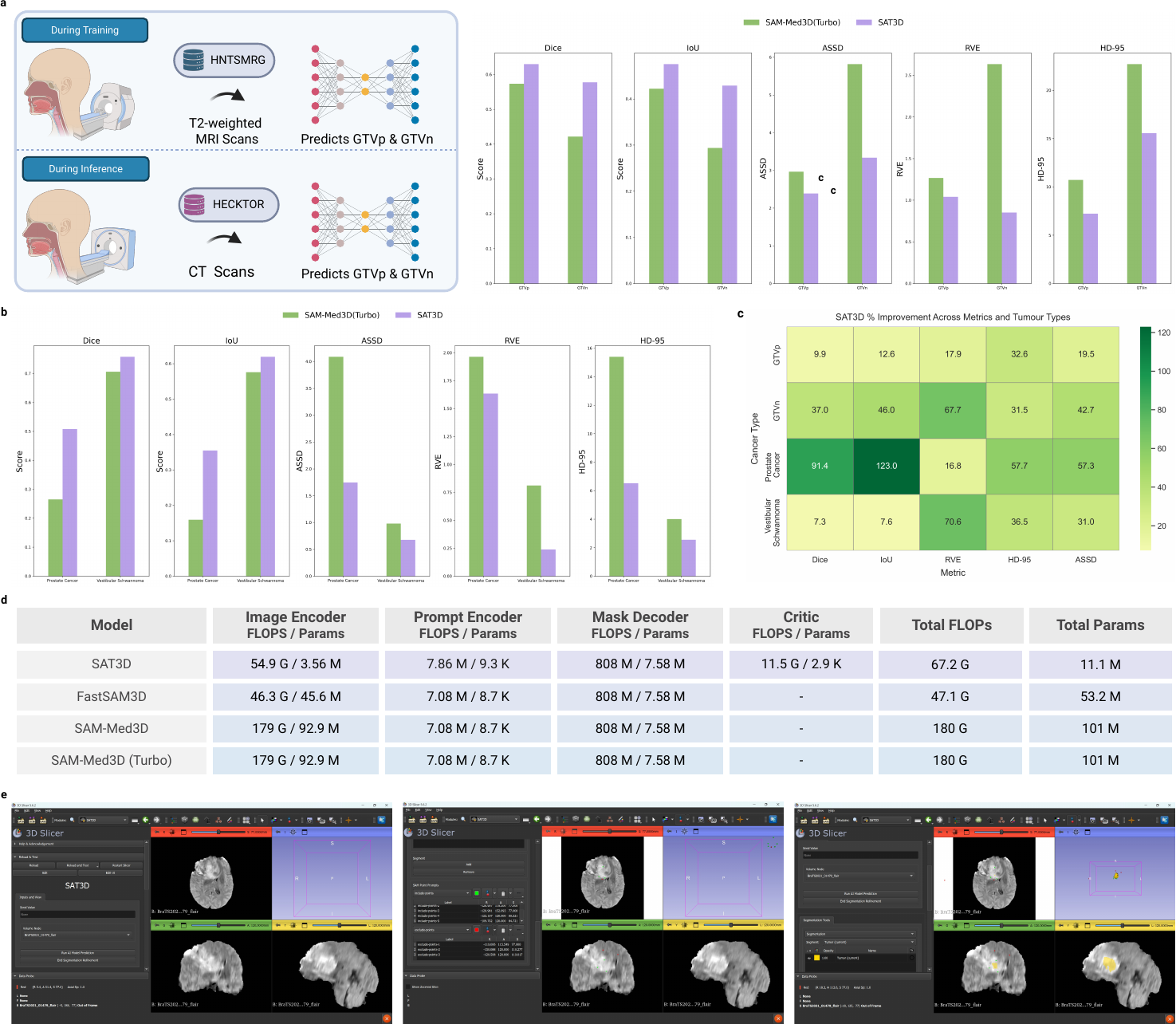}
    \caption{\scriptsize \textbf{Generalizability analysis.} \textbf{a.} Quantitative performance comparison on out-of-distribution head and neck tumour segmentation (HECKTOR 2022) for primary (GTVp) and nodal (GTVn) regions. Although GTVp and GTVn classes were included during training on MRI (HNTSMRG 2024), the model was not exposed to the CT modality, demonstrating cross-modality transfer.  \textbf{b.} Zero-shot segmentation performance results on unseen prostate cancer (Prostate158) and vestibular schwannoma (CrossMoDa 2022) datasets, illustrating cross-organ and cross-modality generalisation.  \textbf{c.} Summary heatmap showing the percentage improvement of SAT3D over SAM-Med3D (Turbo) across metrics and tumour types for out-of-distribution data (See Extended Data Fig. 3 for Qualitative Comparison). \textbf{d. } Computational complexity analysis of SAT3D and competing VFMs. The table reports the number of floating-point operations (FLOPs) and trainable parameters (Params) for each major architectural component under an input size of 128 $\times$ 128 $\times$ 128. \textbf{e. } This visualisation reel showcases the step-by-step refinement of a segmentation mask for a medical scan using the SAT3D Slicer plugin. In this process, clinicians or experts can provide point prompts based on their expertise, while the SAT3D model generates and iteratively refines the segmentation mask guided by these prompts.}
    \label{fig:figure_5}
\end{figure*}

\paragraph{Zero-shot and Out-of-distribution Generalisation.}
To evaluate the robustness and generalisability of SAT3D beyond its training distribution, we performed extensive zero-shot and cross-domain inference across three representative tasks: (i) head and neck tumours using the HECKTOR 2022 dataset, focusing on primary (GTVp) and nodal (GTVn) tumour segmentation; (ii) prostate cancer using the Prostate158 dataset~\cite{adams2022prostate158} with T2-weighted MRI; and (iii) vestibular schwannoma segmentation using the Cross-Modality Domain Adaptation (CrossMoDa) Challenge 2022 dataset~\cite{dorent2023crossmoda,wijethilake2025crossmoda} with contrast-enhanced T1-weighted MRI.
SAT3D was trained on the HNTSMRG 2024 dataset, which includes T2-weighted MRI scans of the head and neck region with annotated GTVp and GTVn labels, along with a diverse set of 17,075 3D volume–mask pairs from 11 publicly available datasets spanning PET, CT, MRI, CTA, and ultrasound modalities. Notably, no samples from the prostate or vestibular regions were seen during training, making these tasks a true assessment of zero-shot generalisation.
As summarised in~\cref{fig:figure_5}, SAT3D consistently outperformed SAM-Med3D (Turbo) across all out-of-distribution evaluation tasks. For head and neck segmentation, SAT3D improved Dice scores from 0.573 to 0.630 (GTVp) and from 0.422 to 0.578 (GTVn), while reducing HD-95 distances by approximately 30\%. On prostate cancer, SAT3D doubled the Dice score (0.507 vs. 0.265) and reduced boundary error (HD-95: 6.52 vs. 15.40 mm), demonstrating strong cross-organ transferability from non-prostate training data. For vestibular schwannoma, SAT3D achieved the highest overall accuracy with a Dice of 0.758 and an ASSD of 0.68 mm, surpassing SAM-Med3D (Turbo) by clear margins across all metrics.
Despite substantial differences in modality, anatomy, and acquisition protocols, SAT3D maintained reliable volumetric and boundary predictions across these unseen domains, highlighting its cross-modality, cross-organ, and cross-task generalisation capability. These findings confirm that SAT3D’s uncertainty-guided prompt learning pipeline enables efficient knowledge transfer, even under severe domain shifts.

\paragraph{Computational Complexity Analysis.}

We benchmarked the computational efficiency of SAT3D against two representative prompt-conditioned 3D foundation models, FastSAM3D and SAM-Med3D, SAM-Med3D (Turbo), under a standardised input of 128 $\times$ 128 $\times$ 128. As summarised in~\cref{fig:figure_5}d, SAT3D achieves a total computational cost of 67.2 G FLOPs with 11.1 M trainable parameters, representing a 2.7$\times$ reduction in computation and a 9$\times$ reduction in parameter count relative to SAM-Med3D (180 G FLOPs, 101 M params). This efficiency is primarily attributed to the lightweight hierarchical image encoder and the parameter-efficient critic module that operates at a lower resolution for uncertainty estimation. In contrast, SAM-Med3D and its Turbo variant rely on heavy transformer backbones, while FastSAM3D, though computationally comparable in FLOPs (47.1 G), exhibits a much larger parameter footprint (53.2 M) due to dense convolutional encoder layers. The block-wise decomposition further highlights that the image encoder dominates the overall cost across all models, whereas the prompt encoder and mask decoder contribute minimally to the total complexity ($<$ 2\%). Overall, SAT3D achieves a superior trade-off between representational capacity and computational efficiency, making it more suitable for real-time clinical deployment and volumetric inference on resource-constrained hardware.

\begin{table*}[ht!]
\centering
\caption{\scriptsize \textbf{Per-block metric values with ranks using the Friedman test across all tumour/cancer types and evaluation metrics.} 
Ranks (in parentheses) indicate the relative performance of each method (1 = best).}
\scriptsize
\resizebox{0.78\linewidth}{!}{
\begin{tabular}{llrrrrr}
\toprule
Metric &             CancerType &  nnUNet (rank) &  SAM-Med3D (rank) &  SAM-Med3D(Turbo) (rank) &  FastSAM3D (rank) &  SAT3D (rank) \\
\midrule
  \multirow{14}{*}{DSC} &          Breast Tumour &   0.4936 (3) &   0.4427 (5) &       0.5954 (2) &   0.4479 (4) &   0.7725 (1) \\
   & Colon Cancer Primaries &   0.5316 (2) &   0.5231 (3) &       0.3965 (5) &   0.4681 (4) &   0.6156 (1) \\
   &                  Edema &   0.8343 (1) &   0.5809 (4) &       0.7045 (3) &   0.3455 (5) &   0.7411 (2) \\
   &       Enhancing Tumour &   0.8287 (1) &   0.5424 (4) &       0.6716 (3) &   0.4875 (5) &   0.7081 (2) \\
   &                   GTVn &   0.7393 (1) &   0.4913 (4) &       0.4986 (3) &   0.4556 (5) &   0.5802 (2) \\
   &                   GTVp &   0.6840 (3) &   0.6475 (4) &       0.7441 (2) &   0.6207 (5) &   0.7946 (1) \\
   &         Hepatic Tumour &   0.7431 (1) &   0.4423 (3) &       0.4164 (4) &   0.3984 (5) &   0.6833 (2) \\
   &          Kidney Tumour &   0.6947 (1) &   0.5330 (4) &       0.6365 (3) &   0.4772 (5) &   0.6806 (2) \\
   &           Liver Tumour &   0.6474 (1) &   0.3350 (4) &       0.3830 (3) &   0.2893 (5) &   0.5446 (2) \\
   &            Lung cancer &   0.7607 (1) &   0.6871 (3) &       0.6077 (5) &   0.6457 (4) &   0.7282 (2) \\
   &  Non-Enhancing Tumour &   0.6822 (1) &   0.5700 (3) &       0.4826 (5) &   0.4920 (4) &   0.6740 (2) \\
   &        Pancreas Cancer &   0.4273 (5) &   0.5949 (3) &       0.5244 (4) &   0.6821 (2) &   0.7036 (1) \\
   &           Renal Tumour &   0.8797 (2) &   0.5625 (4) &       0.9038 (1) &   0.4804 (5) &   0.8601 (3) \\
   &                 Tumour &   0.5180 (1) &   0.0951 (5) &       0.1392 (3) &   0.1047 (4) &   0.3256 (2) \\ \midrule
\multirow{14}{*}{IoU} &          Breast Tumour &   0.3738 (3) &   0.3030 (5) &       0.4378 (2) &   0.3138 (4)&   0.6329 (1) \\
    & Colon Cancer Primaries &   0.4479 (2) &   0.3678 (3) &       0.2660 (5) &   0.3139 (4) &   0.4640 (1) \\
    &                  Edema &   0.7531 (1) &   0.4273 (4) &       0.5683 (3) &   0.2192 (5) &   0.6104 (2) \\
    &       Enhancing Tumour &   0.7658 (1) &   0.3942 (4) &       0.5589 (3) &   0.3407 (5) &   0.5844 (2) \\
    &                   GTVn &   0.6542 (1) &   0.3713 (4) &       0.3900 (3) &   0.3427 (5) &   0.4792 (2) \\
    &                   GTVp &   0.5864 (3) &   0.4879 (4) &       0.6028 (2) &   0.4545 (5) &   0.6691 (1) \\
    &         Hepatic Tumour &   0.6237 (1) &   0.3151 (3) &       0.2986 (4) &   0.2811 (5) &   0.5387 (2)\\
    &          Kidney Tumour &   0.6009 (1) &   0.3971 (4) &       0.5201 (3) &   0.3576 (5)&   0.5581 (2) \\
    &           Liver Tumour &   0.5314 (1) &   0.2398 (4) &       0.3051 (3)&   0.2049 (5) &   0.4209 (2) \\
    &            Lung cancer &   0.6261 (1) &   0.5275 (3)&       0.4680 (5) &   0.4902 (4) &   0.5789 (2) \\
    &  Non-Enhancing Tumour &   0.6058 (1) &   0.4318 (3) &       0.3554 (4) &   0.3457 (5) &   0.5462 (2) \\
    &        Pancreas Cancer &   0.3428 (5) &   0.4463 (3) &       0.3784 (4) &   0.5259 (2) &   0.5594 (1) \\
    &           Renal Tumour &   0.8038 (2)&   0.4107 (4) &       0.8269 (1) &   0.3413 (5) &   0.7590 (3) \\
    &                 Tumour &   0.4010 (1) &   0.0562 (5) &       0.0846 (3) &   0.0600 (4) &   0.2217 (2) \\ \midrule
    \multirow{14}{*}{RVE} &          Breast Tumour &   1.1122 (5) &   0.5138 (4) &       0.4010 (2) &   0.4937 (3)&   0.2029 (1) \\
    & Colon Cancer Primaries &   0.4211 (2) &   0.4978 (4) &       1.5046 (5) &   0.1282 (1) &   0.4733 (3) \\
    &                  Edema &   0.3738 (1) &   0.8922 (5) &       0.6806 (4) &   0.5247 (3) &   0.3783 (2) \\
    &       Enhancing Tumour &   0.2116 (1)&   9.5892 (4) &      22.0469 (5) &   4.7139 (2) &   4.7267 (3) \\
    &                   GTVn &   0.2547 (1)&   0.5031 (4) &       0.9225 (5) &   0.4998 (3) &   0.4633 (2) \\
    &                   GTVp &   0.4247 (5) &   0.2923 (3) &       0.2765 (2) &   0.3251 (4) &   0.2371 (1) \\
    &         Hepatic Tumour &   0.3196 (1) &   1.1268 (4) &      10.9299 (5) &   0.8048 (3)&   0.4601 (2) \\
    &          Kidney Tumour &   1.0470 (5) &   0.9916 (4) &       0.5228 (2) &   0.5287 (3) &   0.4598 (1) \\
    &           Liver Tumour &   0.6024 (1) &   3.9138 (4) &      29.0444 (5) &   0.7835 (2)&   0.9504 (3) \\
    &            Lung cancer &   0.2631 (3) &   0.2473 (2) &       1.2716 (5) &   0.2284 (1) &   0.2675 (4) \\
    &  Non-Enhancing Tumour &   1.0062 (1) &   7.5343 (4) &       8.4825 (5) &   4.4375 (3) &   4.0314 (2) \\
    &        Pancreas Cancer &   0.6468 (2) &   1.8522 (4) &       2.5067 (5) &   0.2343 (1) &   0.7527 (3) \\
    &           Renal Tumour &   0.2563 (3) &   0.7572 (5) &       0.1191 (1) &   0.5929 (4)&   0.1739 (2) \\
    &                 Tumour &   0.9502 (1) &  18.8196 (5) &      15.8959 (4) &  13.4983 (3) &   1.0713 (2) \\ \midrule
  \multirow{14}{*}{HD-95} &          Breast Tumour & 183.1132 (5) &  27.2437 (3) &      19.2679 (2) &  31.3763 (4) &  12.5386 (1) \\
  & Colon Cancer Primaries &  30.7813 (5) &  13.0452 (2) &      23.7217 (4) &  16.0763 (3) &   8.1746 (1) \\
 &                  Edema &   5.3198 (1) &   9.3849 (4) &       7.8971 (3) &  18.3947 (5) &   6.0057 (2) \\
  &       Enhancing Tumour &   5.6289 (1) &   9.6938 (4) &       7.5662 (3) &   9.9828 (5) &   5.9685 (2) \\
  &                   GTVn &  10.2727 (1) &  20.6621 (3) &      17.5299 (2) &  24.4360 (5) &  23.3643 (4) \\
  &                   GTVp &   6.4803 (2) &   8.1313 (5) &       6.5859 (3) &   7.2744 (4) &   4.8969 (1) \\
  &         Hepatic Tumour &  33.6824 (5) &  29.6610 (2) &      33.5505 (4) &  31.1887 (3) &  22.9848 (1) \\
  &          Kidney Tumour &  48.6013 (5) &  20.2061 (3) &      18.9145 (2) &  25.1649 (4) &  16.0699 (1) \\
  &           Liver Tumour &  39.3002 (3) &  36.0651 (2) &      33.8450 (1) &  58.7444 (5) &  47.3065 (4) \\
  &            Lung cancer &  40.1860 (5) &   5.0946 (2) &       9.2592 (4) &   6.1915 (3) &   4.9612 (1) \\
  &  Non-Enhancing Tumour &   9.8704 (5) &   7.5905 (2) &       8.3090 (3) &   9.4227 (4) &   6.2568 (1) \\
  &        Pancreas Cancer &  11.7032 (4) &   6.5129 (3) &      12.3708 (5) &   5.8101 (2) &   4.9051 (1) \\
  &           Renal Tumour &   9.3536 (3) &  14.1565 (5) &       4.2674 (2) &  12.1761 (4) &   4.1357 (1) \\
  &                 Tumour & 146.9472 (3) & 166.8653 (4) &     143.9810 (2) & 181.7408 (5) & 142.8242 (1) \\ \midrule
  \multirow{14}{*}{ASSD} &          Breast Tumour &  33.7381 (5) &   6.2433 (3)&       4.2877 (2) &   7.9106 (4) &   2.2182 (1) \\
   & Colon Cancer Primaries &  20.3757 (5) &   5.1538 (2) &       9.1906 (4)&   6.7744 (3) &   2.3654 (1) \\
   &                  Edema &   1.2962 (1) &   2.7100 (4) &       1.6124 (3) &   5.1871 (5) &   1.3274 (2) \\
   &       Enhancing Tumour &   1.1110 (1) &   2.7536 (5) &       1.8350 (3) &   2.7250 (4) &   1.4323 (2) \\
   &                   GTVn &   1.3781 (1) &   6.7136 (3)&       6.9476 (4) &   9.9814 (5) &   6.1626 (2) \\
   &                   GTVp &   1.7220 (3) &   2.3483 (5) &       1.6694 (2) &   2.3248 (4) &   1.2823 (1) \\
   &         Hepatic Tumour &   6.7348 (2) &   9.1875 (3) &      10.4830 (4) &  12.6373 (5) &   4.9643 (1) \\
   &          Kidney Tumour &  11.6181 (5) &   8.0601 (3) &       6.6552 (2) &  10.7287 (4) &   4.8076 (1) \\
   &           Liver Tumour &  11.3808 (2) &  12.9929 (4) &      11.9603 (3) &  19.1599 (5) &   8.1450 (1) \\
   &            Lung cancer &   7.7585 (5)&   1.4980 (2) &       2.7793 (4) &   1.8608 (3)&   1.2933 (1) \\
   &  Non-Enhancing Tumour &   2.8862 (4) &   2.2835 (2) &       3.0033 (5) &   2.7851 (3) &   1.5564 (1) \\
   &        Pancreas Cancer &   4.5737 (5) &   2.5817 (3) &       3.5641 (4) &   1.6613 (2)&   1.6194 (1) \\
   &           Renal Tumour &   2.0667 (3) &   4.5570 (4) &       0.9772 (1)&   5.2250 (5) &   1.3309 (2) \\
   &                 Tumour &  41.1325 (2)&  58.0407 (5) &      47.2259 (3) &  54.4771 (4) &  31.9899 (1) \\
  \bottomrule
\toprule
  \multicolumn{7}{c}{\textbf{Friedman Test Summary - Friedman $\chi^2$ ($p$-value) : 89.54 ($1.65 \times 10^{-18}$)}} \\
\midrule
\multicolumn{2}{c}{\textbf{Average Rank (Overall Order)}} & 2.46 (2) & 3.63 (4) & 3.29 (3) & 3.90 (5) & 1.73 (1) \\ 
 \bottomrule 
  \end{tabular}
  }
  \label{tab:table_1}
\end{table*}

\paragraph{Interactive Demonstration via 3D-slicer integration.}
To facilitate clinical translation and interactive exploration, we developed a dedicated 3D Slicer plugin that integrates SAT3D for volumetric tumour segmentation and visualisation as shown in~\cref{fig:figure_5}e~\cite{pieper20043d,shen2024fastsam}.
The plugin enables users to load patient scans directly into the 3D Slicer environment, automatically preprocess the data, and perform inference using the trained SAT3D model, all without requiring command-line execution.
Predicted segmentations are rendered in real time within Slicer’s 3D viewer, supporting overlay visualisation, threshold-based refinement, and region-of-interest editing.
This deployment demonstrates the practical applicability of SAT3D beyond research settings, bridging model inference with clinical imaging workflows and facilitating reproducible evaluation across institutions.
In our source code, we provide details on how to add this plugin to the Slicer software, as well as the source code of the Slicer plugin. 

\section*{Discussion}
In this work, we present SAT3D, an uncertainty-aware, prompt-driven foundation model designed for robust 3D tumour and lesion segmentation across a broad range of medical imaging modalities and anatomical regions. SAT3D builds upon recent methods in SAM architectures, extending them to the volumetric medical domain through 3D adaptations, uncertainty integration, and training on diverse tumour and cancer-centric datasets.

Our results demonstrate that SAT3D achieves significantly improved segmentation accuracy and generalisability across a variety of challenging tumour types, including brain tumours, head and neck primaries, and abdominal cancers, while maintaining generalisability across multiple imaging modalities, including CT, MRI (T1-w, T2-w, FLAIR, contrast-enhanced T1-w), PET, CTA, and ultrasound. Notably, SAT3D performs competitively when benchmarked against task-specific state-of-the-art models, such as nnUNet, and surpasses existing prompt-based foundation models, including SAM-Med3D, SAM-Med3D (Turbo), and FastSAM3D, especially in cross-domain and out-of-distribution evaluation settings.

A key feature of SAT3D is its hybrid prompt mechanism, which leverages both sparse (point-based) and dense (mask-based) cues, dynamically updated during training. By incorporating uncertainty estimates into the prompting and prediction loop, the model is better equipped to handle ambiguous boundaries, low-contrast regions, and heterogeneous lesion presentations—scenarios that are typically challenging for traditional segmentation methods. This property is particularly important in clinical oncology, where consistency and reliability in segmentation are critical for downstream tasks such as treatment planning and disease monitoring.

Beyond segmentation performance, SAT3D supports volumetric biomarker extraction, such as tumour volume and lesion spread, enabling integration into clinical workflows for assessing disease burden and treatment response. Its foundation-style architecture allows adaptation to new segmentation targets with minimal retraining, positioning SAT3D as a scalable solution for multi-organ, multi-modality clinical applications.

From a computational complexity perspective, SAT3D demonstrates an efficient balance between architectural expressiveness and computational cost. Despite integrating an uncertainty critic (~11.5 GFLOPs), the overall footprint remains comparable to or lower than existing volumetric foundation models such as SAM-Med3D, achieving superior accuracy at similar computational budgets. This highlights SAT3D’s efficiency and scalability for large-volume inference without sacrificing precision.

Another important consideration is the handling of multi-modal imaging datasets, where multiple imaging modalities, such as PET-CT or various MRI sequences combined, serve as input for producing predictions. While task-specific architectures such as nnUNet are trained with all modalities jointly as multi-channel inputs, VFMs, including SAT3D, were trained with single-modality loaders. This limits their ability to directly leverage complementary contrasts in datasets like AutoPET 2024 or BraTS 2021, where combining modalities provides a richer signal for segmentation. As a result, nnUNet retains an advantage in settings where all modalities are consistently available. On the other hand, VFMs provide a unique strength in scenarios where only a single modality is available. In such cases, nnUNet models trained with the expectation of multi-modal input often fail to generalise, whereas SAT3D can still generate reliable predictions from whichever contrast is provided (\eg, T1-w or FLAIR alone). This flexibility is particularly valuable in clinical environments where incomplete imaging protocols, missing contrasts, or modality restrictions are common, as well as in retrospective studies where only a subset of scans may be available.

Nonetheless, several other limitations remain. First, while our dataset encompasses a wide range of modalities and tumour types, some modalities (\eg, ultrasound) and less common anatomical sites are still underrepresented. Second, like other prompt-based models, SAT3D’s performance is influenced by prompt quality; poorly localised or ambiguous prompts can degrade accuracy, especially in edge cases with diffuse tumour margins. Lastly, inference time and memory consumption remain higher than conventional 2D models, due to the 3D architecture and prompt embedding pipeline.

In our future works, we aim to extend SAT3D in several directions. First, we plan to combine information from multiple imaging modalities (\eg, CT and PET) to leverage their complementary features, rather than training each modality separately. Second, we will explore using the uncertainty maps produced by the critic to guide test-time adaptation and active learning, allowing the model to refine its predictions automatically and highlight cases that need expert review. Third, we will focus on improving model efficiency by reducing memory and computation requirements, making SAT3D faster and more practical for real-time clinical use. We also plan to expand the dataset to include more underrepresented modalities, such as ultrasound and less common tumour sites. Finally, our current 3D Slicer plugin serves as a preliminary design for interactive visualisation and prompt-based refinement. In future iterations, we aim to extend this interface for seamless integration with picture archiving and communication systems (PACS), enabling clinicians to provide feedback, interact with the model’s prompts, and support continuous model improvement within real-world clinical workflows.

In conclusion, SAT3D advances the field of foundation models in medical image segmentation by introducing a reliable, prompt-driven framework tailored for the analysis of complex, volumetric tumours. Its generalisation capabilities and uncertainty-aware design represent a step forward toward building clinically adaptable, domain-agnostic segmentation systems that reduce the need for task-specific model engineering and manual annotation.

\section*{Method}
We begin this section by providing methodological details of the SAT3D model architecture (\cref{fig:figure_1}d presents the schematic diagram). We then provide details of the architectural components and the training pipeline, including the objective function.

\paragraph{Preliminary.}
In this study, we closely align with the blueprint of the SAM's architecture, leveraging its core components to refine its functionality for semi-supervised MIS. SAM's structure comprises three elements.
In the \textbf{Image Encoder}, SAM utilises an MAE~\cite{he2022masked} pre-trained Vision Transformer (ViT)~\cite{dosovitskiy2020image} to extract features.
This component uses 2D patch embeddings and learnable positional encodings to convert the input image into image embeddings.
The \textbf{Prompt Encoder} module handles both sparse prompts (\ie, points or boxes) and dense (masks) prompts. Sparse prompts are encoded using fixed absolute positional encodings, then merged with learned embeddings tailored to each prompt type. Dense prompts, conversely, undergo encoding through a convolution to generate dense prompt embeddings. 
Employing a lightweight \textbf{Mask Decoder}, SAM efficiently maps image embedding along with a set of prompt embeddings to produce an output mask. Each transformer layer comprises four steps: self-attention on tokens, cross-attention between tokens and the image embedding, token updates via a point-wise Multi-Layer Perceptron (MLP), and cross-attention that updates the image embedding with prompt details. Following processing through the transformer layers, the feature map undergoes up-sampling and is converted into segmentation masks using an MLP. Notably, all transformer layers capture 2D geometric information during the forward pass.

\paragraph{Problem Formulation \& Notation.}
Throughout our paper, we denote vectors and matrices in bold lowercase $\mathbf{x}$ and bold uppercase $\mathbf{X}$, respectively. The norm of a vector is denoted by $\| \cdot \|$ and $\|\mathbf{x}\|_1 = \sum_i \bigl\lvert\mathbf{x}[i] \bigl\lvert$, where $\mathbf{x}[i]$ denotes the element at position $i$ in $\mathbf{x}$. The inner product between vectors is represented by $\langle \cdot,\cdot \rangle$ and $\|\mathbf{x}\|_2^2 = \langle \mathbf{x},\mathbf{x} \rangle$. When norms and inner products are used over 3D tensors, we assume that the tensors are flattened accordingly. For example, for 3D tensors $\mathbf{A}$ and $\mathbf{B}$, $\langle \mathbf{A},\mathbf{B} \rangle = \sum_{i,j,k} \mathbf{A}[i,j,k]\mathbf{B}[i,j,k]$. 

Let  $\mathcal{X} = \{(\vec{X}_i,\vec{Y}_i)\}_{i=1}^n$ be a labelled set with $n$ samples, where each sample $(\vec{X}_i,\vec{Y}_i)$ consists of an volume $\vec{X}_i \in \mathbb{R}^{C \times H \times W \times D}$ and its associated ground-truth segmentation mask (binary mask) $\vec{Y}_i \in \{0,1\}^{1 \times H \times W \times D}$.
Here, $C$, $H$, $W$, and $D$ represent the number of channels, height, width, and depth of the input medical volume.

\paragraph{Proposed Network \& Baselines.}
Building upon the building blocks of SAM and its adaptations~\cite{SAM,SAMMED2D,SAMMED3D}, similar to SAM's structure, SAT3D comprises three main components in the segmentation pipeline (See~\cref{fig:figure_1}d): A 3D Image Encoder, a 3D Prompt Encoder, and a 3D Mask Decoder. Additionally, based on our previous works~\cite{peiris2021duo,peiris2023uncertainty}, we incorporate a discriminator/critic with a CNN decoder to guide prompt generation by measuring uncertainty through confidence maps, which serve as both a surrogate model and a subjective referee. 
When the discriminator is applied to an input (\eg, segmentation mask from Mask decoder), it does not just output a single scalar value (real or fake). Instead, it produces a confidence map that provides a voxel-wise~\cite{cirillo2020vox2vox}. Each value in this map represents the critic's confidence that the corresponding region in the input (prediction generated from the model) is real (or fake). 

\noindent \textbf{3D Image Encoder:}
Here, we employ a design that combines 3D patch partitioning with a hierarchical vision transformer architecture based on Shifted Windows (SWIN) transformer blocks and 3D patch merging layers to enable efficient volumetric feature extraction from medical scans~\cite{liu2021swin}. Initially, the 3D patch partitioning layer divides the input volume into non-overlapping 3D patches, each processed by a linear embedding layer to produce token representations with an embedding dimension of 48, corresponding to the tiny variant of the Swin Transformer. These token embeddings are then passed through stacked SWIN transformer blocks that integrate window-based multi-head self-attention (MSA) and shifted window MSA (SW-MSA) to capture both local and global spatial relationships directly~\cite{peiris2021volumetric}. Unlike the large variants of vision transformer architectures~\cite{dosovitskiy2020image,liu2021swin}, which tend to be computationally demanding during inference~\cite{engelmann2025training}, this compact configuration provides a balanced trade-off between representational power and efficiency, making it well-suited for 3D medical segmentation tasks.

\noindent \textbf{3D Prompt Encoder:}
In our approach, we utilise both sparse and dense prompts to guide the segmentation process. Sparse prompts are generated using the available ground truth annotations and previous predictions. In the initial iteration, no point prompts are assumed, and thus, sparse prompts are not provided. From the second iteration onward, both ground truth and prior prediction masks are available, enabling automatic generation of sparse prompts during training. Dense prompts are derived from the previous predictions and the confidence maps generated by the critic for those predictions. Similar to sparse prompting, the first dense prompt is initialised with a blank (all-zero) mask. Using both prompt types during the training, we compute the corresponding prompt embeddings, which are then passed to the 3D Mask Decoder to produce the segmentation output.
In contrast to other SAM-based VFMs for medical imaging, we introduce confidence maps as auxiliary dense prompts to further guide the model in uncertain regions (specifically on boundaries) of the previous prediction produced by the model.
To integrate the confidence or uncertain region information provided by the critic into the segmentation process, we designed an embedding strategy that jointly encodes the predicted masks and their associated confidence maps. Specifically, given a predicted mask and its corresponding confidence map, we applied separate downscaling modules to project them into a lower-dimensional feature space.
The resulting embeddings were then concatenated along the channel dimension to form a joint dense representation, which served as the input for subsequent processing. This design ensures that both the spatial structure of the masks and the critic-derived confidence information are preserved and fused in the embedding space.

\noindent \textbf{3D Mask Decoder:}
The 3D Mask Decoder reconstructs dense segmentation maps from encoded feature embeddings using a hierarchical attention–based decoding process.
It employs a Two-Way Transformer to fuse image and prompt embeddings, enabling contextual interaction between query tokens and volumetric features. The resulting feature representations are then progressively upscaled through 3D transposed convolutions, restoring spatial resolution while preserving fine structural detail.

\noindent \textbf{Critic Network:}
With the introduction of Generative Adversarial Learning by Goodfellow~\etal, various tasks have been explored to examine models' generative ability and discriminative features~\cite{goodfellow2014generative}. 
In the Generative Adversarial Network (GAN) training, the Generator and the Discriminator play a two-player game to generate realistic, high-quality predictions. 
Since quality plays a significant role in the medical AI domain due to its variability and inherent uncertainty, we introduced a discriminator, which we refer to as a critic network, into the SAT3D training pipeline. This network performs voxel-wise binary classification on the samples generated by determining whether they are real or fake images. If the samples are classified as real images, the critic labels them as one, while fake images are labelled zero.

\paragraph{Training Objective.}
As illustrated in~\cref{fig:figure_1}d, SAT3D comprises a segmentation network (denoted by $\mathcal{F}(\cdot)$, which is a functional composition of 3D image encoder, 3D prompt encoder, and 3D mask decoder) and a critic network (denoted by $\psi(\cdot)$), characterized by parameters denoted as $\vtheta_G$ and $\vtheta_C$, respectively. 
Here, $\vtheta_G$ is the aggregation of network parameters of 3D Image encoder ($\vtheta_E$), 3D Mask decoder ($\vtheta_M$) and 3D Prompt encoder ($\vtheta_P$).
In the context of training the SAT3D segmentation model, inspired by recent works~\cite{peiris2021duo,peiris2023uncertainty,peiris2022reciprocal}, we propose to optimize the following min-max problem:
\begin{equation}
    \label{eqn:min_max_overall_loss} 
    \min_{\vtheta_\text{G}}\max_{\vtheta_\text{C}} ~\Loss(\vec{\Theta}; \mathcal{X}) \;.
\end{equation}
Here, ${\vtheta}_G$ encompasses all the parameters of sub-networks of the segmentation model, \ie, $\vtheta_E$, $\vtheta_M$, $\vtheta_P$ and $\vec{\Theta}$ encompasses all the parameters of the network pipeline, \ie, ${\vtheta}_G,{\vtheta}_C$.The aforementioned min-max problem in Equation~\ref{eqn:min_max_overall_loss} is designed to assess whether the prediction masks generated by segmentation networks belong to the same or different distribution compared to the ground truth distribution.

\noindent \textbf{Loss function formulation:}
As depicted in~\cref{fig:figure_1}d, the image embeddings for $\mathcal{X}$ are derived through a SWIN Transformer block-based 3D encoder network. In SAT3D, we use the SWIN version with an embedding dimension of 48. The critic network also produces auxiliary masks based on confidence maps derived from the previous predictions.
To train the SAT3D, we jointly minimize the Dice Loss ($\Loss_{\text{dice}}$) and Cross-Entropy (CE) loss ($\Loss_{\text{ce}}$) (computed voxel-wise) ($\Loss_{\mathrm{s}}$) together with adversarial loss ($\Loss_{\mathrm{c}}$) and uncertainty loss ($\Loss_{\mathrm{u}}$). Our multi-task loss function is defined as:
\begin{equation}
    \label{eqn:gen_loss}
    \Loss(\vtheta_\text{G}; \mathcal{X}) \coloneqq 
    \Loss_{\mathrm{s}}({\vtheta}_\text{G}; \mathcal{X}) 
    + \lambda_c \Loss_{\mathrm{c}}({\vtheta}_\text{G}; \mathcal{X}) + \lambda_u \Loss_{\mathrm{u}}({\vtheta}_\text{G}; \mathcal{X}) \;, 
\end{equation}
where $\lambda_c$ and $\lambda_u$ are weights to control loss contributions during training. We set $\lambda_{c} = 0.01$ and $\lambda_{u} = 0.1$ in all our experiments. 

At each step $t$ in this iterative process, given a medical volume ($\mathbf{X}_i$) and its ground truth ($\mathbf{Y}_i$), we perform an iterative refinement to generate plausible segmentation masks. Specifically, suppose the predefined number of prompts is set to $m=5$, meaning that the model generates five predictions in each training step. In that case, we will obtain five valid masks for a single medical volume or patient case during each iteration of training.
In the first step, the previous mask is empty. Then, at each subsequent step, the previous prediction $\hat{\mathbf{Y}}_{i}^{t-1}$ and its confidence map $\mathbf{Z}_{i-1}^{t-1}$ (binarised mask generated by thresholding the critic's uncertainty information) are used to dense prompts. 
We learn $\mathcal{F}(\vtheta_\text{G};\mathbf{X}_i; m; \hat{\mathbf{Y}}_{i-1}^{t-1}; \mathbf{Z}_{i-1}^{t-1})$ (later in the manuscript we use the simplified term $\mathcal{F}(\vtheta_\text{G};\mathbf{X}; m; \hat{\mathbf{Y}}^{t-1}; \mathbf{Z}^{t-1})$) that produces $t^{th}$ (final valid mask or predicted segmentation mask at step $t$) segmentation $\hat{\mathbf{Y}}_{i}^{t}$ (or $\hat{\mathbf{Y}}^{t}$) and has learnable parameters $\vtheta_G$. 
The primary segmentation loss is defined as:
\begin{equation}
\label{eqn:sup_loss}
     \Loss_{\mathrm{s}}({\vtheta}_\text{G}; \mathcal{X}) = \Loss_{\text{ce}}(\vec{\vtheta}_\text{G};\mathcal{X}) + \Loss_{\text{dice}}(\vtheta_\text{G};\mathcal{X}),
\end{equation}
In our training pipeline, we use a critic network which has the functionality of $\psi:[0,1]^{H \times W \times D} \to [0,1]^{H \times W \times D}$ that helps the segmentation network to generate realistic segmentation masks using min-max game as defined in~\cref{eqn:min_max_overall_loss}. The adversarial loss for the training segmentation network is defined as:
\begin{multline}
    \Loss_{\text{c}}(\vtheta_\text{G}; \mathcal{D}) \coloneqq -\mathbb{E}_{(\vec{X},\vec{Y} \sim \mathcal{D})} 
    \Big[\sum_{\substack{a , b , c }} 
    \mathbf{1}\log\big(\psi(\mathcal{F}(\vec{X}, m, \hat{\vec{Y}}^{t-1}, \vec{Z}^{t-1}))[a,b,c]\big) 
    \Big]\;,
    \label{eqn:loss_gen_adv}
\end{multline}
As discussed, our approach leverages previous predictions and their associated confidence maps to generate subsequent segmentation masks. To ensure the accuracy of these confidence maps and provide guidance for creating valid prompts, we integrate a spatial masked CE loss to train the model based on the notion of uncertainty~\cite{peiris2023uncertainty}. Here, we make the masked confidence map by binarising the uncertainty information using a predefined threshold of $T=0.3$. 
This enables the critic to identify confident regions within the predictions generated by the network. Therefore, the masked loss is calculated for labelled data and defined as follows:
\begin{multline}
    \Loss_{\text{u}}(\vtheta_\text{G}; \mathcal{X}) \coloneqq -\mathbb{E}_{(\vec{X},\vec{Y} \sim \mathcal{X})} \Big[
    \sum_{\substack{a , b , c }} 
    \mathbf{1}\big( \psi(\mathcal{F}(\vec{X}, m, \hat{\vec{Y}}^{t-1}, \vec{Z}^{t-1})[a,b,c] > \text{T} \big) \\
    \vec{Y}[a,b,c] 
    \log \big(\mathcal{F}(\vec{X}, m, \hat{\vec{Y}}^{t-1}, \vec{Z}^{t-1})[a,b,c] \big) 
   \Big]\;.
    \label{eqn:loss_mask}
\end{multline}
We use predictions and ground truth labels to train the critic network. We define the adversarial loss as maximising the log-likelihood as:
\begin{multline}
    \Loss_{\text{d}}(\vtheta_\text{C}; \mathcal{X}) \coloneqq \mathbb{E}_{(\vec{X}, \vec{Y}) \sim \mathcal{X}} \bigg[
    \sum_{\substack{a , b , c }} 
    \mathbf{1}
    \Big\{
    \eta \log\big(  \psi(\vec{Y})[a,b,c]\big) 
    + (1-\eta)\log\big(1 - \psi_i(\mathcal{F}(\vec{X}, m, 
     \\ 
    \hat{\vec{Y}}^{t-1}, \vec{Z}^{t-1}))[a,b,c]\big) 
    \Big\}\bigg]\;,
    \label{eqn:loss_critic}
\end{multline}
where $\eta = 0$ when the sample is a prediction mask from a segmentation network, and $\eta = 1$ when the sample is obtained from the ground truth label distribution.

\paragraph{Training \& Testing Datasets.}
To build a foundation model with strong generalisation capabilities for unseen tasks, we train our model on a large-scale, diverse collection of medical images sourced from publicly available online datasets. Our data encompasses 11 different datasets, covering various medical domains and imaging modalities (\ie, CT, MRI, FDG-PET, Ultrasound, CTA). These datasets include images of a wide array of organs, such as the brain, head and neck, breast, lungs, abdomen, and whole body. A detailed list is shown in~\cref{fig:figure_2}a, and these datasets will be provided along with our code.
In this study, we divided the 11 publicly available datasets into two parts: a training split (85\%), and a testing split (15\%) as shown in~\cref{fig:figure_2}a. All the performance comparisons are based on the test dataset. We evaluated all methods, including our own, on the mentioned datasets during the inference process. During our evaluation of model robustness and generalisability analysis, we utilised the HECKTOR 2022 training dataset's CT scans~\cite{andrearczyk2022overview}, the prostate158 dataset's T2-weighted MRI scans~\cite{adams2022prostate158}, and the CrossMoDa 2022 dataset's contrast-enhanced T1-weighted MRI scans~\cite{dorent2023crossmoda,wijethilake2025crossmoda}.
We benchmark three prompt-conditioned vision foundation models: SAM-Med3D, SAM-Med3D (Turbo), and FastSAM3D, and one fully supervised volumetric baseline, nnUNet. To ensure fair comparison, all methods are evaluated under the same task definition and metrics. Prompt-based models are tested with a fixed number of $K \in \{5, 10, 15, 20\}$ foreground points per case, while the supervised baseline operates without prompts. If a method produces multiple candidate masks for a given number of points, we report, for each case, the candidate achieving the highest Dice coefficient, representing an upper performance bound. Only foreground points are used; no background points, boxes, or scribbles are applied. Details about comparison methods, full implementation details, dataset preparation steps, and training configurations are provided in the Supplementary Information.

\paragraph{Data pre-processing \& augmentation.}  
A total of 9,945 3D scans were collected, yielding 20,103 image-mask pairs after converting tumour and cancer primary regions into binary segmentation targets. All volumes were cropped or padded to a fixed size of \(128 \times 128 \times 128\) voxels and normalised using z-score normalisation during the model training. To improve model robustness and generalisation, random rotations and random flips along the three spatial axes were applied during training. The sliding window approach has been used during inference to produce a prediction for the entire volume, where the patch size of \(128 \times 128 \times 128\) is considered.

\paragraph{Training details.}  
The proposed SAT3D model was implemented in PyTorch and trained on two NVIDIA A6000 GPUs (48 GB of memory). Ablations were trained using two Setonix GPUs (128 GB of memory) at the Pawsey Supercomputing Centre, enabling validation of the training pipeline on both NVIDIA and AMD hardware platforms.  
All experiments used a SWIN-based 3D image encoder, prompt encoder, mask decoder and an uncertainty-aware critic. Input volumes were aligned to canonical orientation, z-normalised within the foreground (voxels~$>$~0), randomly flipped along spatial axes, and cropped or padded to \(128^3\) voxels.  
The model was trained using the AdamW optimizer (learning rate~=~\(8\times10^{-4}\), weight decay~=~\(1\times10^{-5}\)) with a cosine annealing learning rate schedule for 500~epochs. Training was performed using automatic mixed precision (AMP), Distributed Data Parallel (DDP) across multiple GPUs, and gradient accumulation of 20 steps with a per-GPU batch size of 3.  
Model checkpoints were saved for the latest, best-loss, and best-Dice states, with the best model selected based on the training loss and Dice score. For the testing phase inference, zero-shot inference was performed on in-distribution and out-of-distribution data (cross-site and cross-target datasets), using the best SAT3D checkpoint as initialisation.

\paragraph{Software Environment and Package Configuration.}
All experiments were conducted using Python 3.9.21 with deep learning models implemented in PyTorch (v2.4.1+cu124), accompanied by Torchvision (v0.19.1+cu124) for supporting utilities. The TorchIO (v0.20.4) library was employed for medical image preprocessing and spatial augmentations, while timm (v1.0.15) provided backbone architectures and model utilities. The surface-distance (v0.1) package was used to compute boundary-based evaluation metrics such as Hausdorff Distance, and mypy (v1.14.1) ensured static type checking and code reliability. For visualisation and clinical integration, a 3D Slicer plugin was developed and tested on the Slicer Desktop version 5.6.4 (Windows), supporting interactive segmentation with SAT3D. The plugin incorporated MONAI’s sliding window inference for efficient patch-wise volumetric prediction, ensuring compatibility with limited GPU memory while maintaining high-resolution segmentation output~\cite{cardoso2022monai}.

\paragraph{Evaluation Procedure.}
The segmentation accuracy of the trained models was evaluated using five standard evaluation metrics: Dice similarity coefficient (DSC), Intersection over Union (IoU), Hausdorff distance (HD), Average Symmetric Surface Distance (ASSD) and Relative Volume Error (RVE)~\cite{taha2015metrics}. The goal is to maximise the DSC and IoU while minimising the HD-95, ASSD and RVE (See the supplementary information for further details).

\section*{Data Availability}
This study incorporates 14 publicly available datasets. Out of these 14 datasets, 11 were used for pretraining, and a subset of them was also employed for evaluation (AutoPET 2024~\cite{gatidis2023autopet,ingrisch_2024_10990932}, HNTSMRG 2024~\cite{wahid_2024_11199559,wahid2024overview}, TDSC-ABUS 2023~\cite{luo2025tumor}, KiPA 2022~\cite{guanyu_yang_2022_6361938}, KiTS 2023~\cite{heller2019kits19,heller2023kits21}, LiTS~\cite{BILIC2023102680}, MSDC Lung~\cite{msd}, MSDC Colon~\cite{msd}, MSDC Pancreas~\cite{msd}, MSDC Hepatic Vessel~\cite{msd}, BraTS 2021~\cite{baid2021rsna,menze2014multimodal,bakas2017advancing,bakas2017segmentation,bakas2017segmentation1}).
The remaining 3 datasets were specifically utilised for zero-shot evaluation and out-of-distribution analysis (HECKTOR 2022~\cite{andrearczyk2022overview}, Prostate158~\cite{adams2022prostate158}, CrossMoDa 2022~\cite{dorent2023crossmoda, wijethilake2025crossmoda}).
All training and validation datasets used in this study are publicly available and can be accessed via the links listed in Supplementary Table 1. All datasets are approved for research use. Additionally, we provide the list of case identification names used for the training and test sets to ensure reproducibility.

\section*{Code Availability}
The code was implemented in Python using the deep learning framework PyTorch\cite{paszke2019pytorch}. The code is publicly available at \href{https://github.com/himashi92/SAT3D}{https://github.com/himashi92/SAT3D}. The source code is provided under the MIT license.
All the pre-trained model weights and supplementary results can be found in the Figshare Project Page \href{https://figshare.com/s/a8c19cd60a57e975390b}{https://doi.org/10.6084/m9.figshare.30155497}.

\section*{Acknowledgements}
This work was supported by the Australian Research Council Discovery Program DP210101863 and Australian Research Council Mid-Career Industry Fellowship IM230100002. 
This work was also supported by resources provided by the Pawsey Supercomputing Centre with funding from the Australian Government and the Government of Western Australia (project No. pawsey1212).

\section*{Author Contributions Statement}
H.P. conceived the initial idea, designed and executed all experiments, prepared benchmark data, conducted all subsequent statistical analyses, and drafted the manuscript.
H.P. developed the core theory, designed the model and the computational framework and analysed the data.
Z.C. directed and supervised the project.
H.P. and Z.C. interpreted the results. 
H.P. and S.W. developed scripts for the final evaluation of the models.
S.W. contributed to the implementation of the segmentation baselines and wrote the supplementary material. 
H.P., Z.C., M.H., S.W., G.E. and M.L. provided scientific insights on the applications and made substantial revisions and edits of the draft manuscript. All the authors read and approved the final manuscript.

\section*{Competing Interests Statement}
None of the authors has any conflicts of interest to report.

\newpage
\clearpage
\section*{Reference}
\hspace*{\fill}

\bibliography{references}

@inproceedings{goodfellow2014generative,
  title={Generative adversarial nets},
  author={Goodfellow, Ian and Pouget-Abadie, Jean and Mirza, Mehdi and Xu, Bing and Warde-Farley, David and Ozair, Sherjil and Courville, Aaron and Bengio, Yoshua},
  booktitle={Advances in neural information processing systems},
  pages={2672--2680},
  year={2014}
}

@inproceedings{ronneberger2015u,
  title={U-net: Convolutional networks for biomedical image segmentation},
  author={Ronneberger, Olaf and Fischer, Philipp and Brox, Thomas},
  booktitle={International Conference on Medical image computing and computer-assisted intervention},
  pages={234--241},
  year={2015},
  organization={Springer}
}

@InProceedings{peiris2022reciprocal,
author="Peiris, Himashi
and Chen, Zhaolin
and Egan, Gary
and Harandi, Mehrtash",
editor="Crimi, Alessandro
and Bakas, Spyridon",
title="Reciprocal Adversarial Learning for Brain Tumor Segmentation: A Solution to BraTS Challenge 2021 Segmentation Task",
booktitle="Brainlesion: Glioma, Multiple Sclerosis, Stroke and Traumatic Brain Injuries",
year="2022",
publisher="Springer International Publishing",
address="Cham",
pages="171--181",
isbn="978-3-031-08999-2"
}

@inproceedings{peiris2021duo,
  title={Duo-SegNet: Adversarial Dual-Views for Semi-Supervised Medical Image Segmentation},
  author={Peiris, Himashi and Chen, Zhaolin and Egan, Gary and Harandi, Mehrtash},
  booktitle={International Conference on Medical Image Computing and Computer-Assisted Intervention},
  pages={428--438},
  year={2021},
  organization={Springer}
}

@article{taha2015metrics,
  title={Metrics for evaluating 3D medical image segmentation: analysis, selection, and tool},
  author={Taha, Abdel Aziz and Hanbury, Allan},
  journal={BMC medical imaging},
  volume={15},
  number={1},
  pages={1--28},
  year={2015},
  publisher={BioMed Central}
}

@article{paszke2019pytorch,
  title={Pytorch: An imperative style, high-performance deep learning library},
  author={Paszke, Adam and Gross, Sam and Massa, Francisco and Lerer, Adam and Bradbury, James and Chanan, Gregory and Killeen, Trevor and Lin, Zeming and Gimelshein, Natalia and Antiga, Luca and others},
  journal={Advances in neural information processing systems},
  volume={32},
  year={2019}
}

@inproceedings{peiris2021volumetric,
  title={A Robust Volumetric Transformer for Accurate 3D Tumor Segmentation},
  author={Peiris, Himashi and Hayat, Munawar and Chen, Zhaolin and Egan, Gary and Harandi, Mehrtash},
  booktitle={International Conference on Medical Image Computing and Computer-Assisted Intervention},
  pages={162--172},
  year={2022},
  organization={Springer}
}

@inproceedings{cirillo2020vox2vox,
  title={Vox2Vox: 3D-GAN for brain tumour segmentation},
  author={Cirillo, Marco Domenico and Abramian, David and Eklund, Anders},
  booktitle={International MICCAI Brainlesion Workshop},
  pages={274--284},
  year={2020},
  organization={Springer}
}

@article{dosovitskiy2020image,
  title={An image is worth 16x16 words: Transformers for image recognition at scale},
  author={Dosovitskiy, Alexey and Beyer, Lucas and Kolesnikov, Alexander and Weissenborn, Dirk and Zhai, Xiaohua and Unterthiner, Thomas and Dehghani, Mostafa and Minderer, Matthias and Heigold, Georg and Gelly, Sylvain and others},
  journal={arXiv preprint arXiv:2010.11929},
  year={2020}
}

@article{baid2021rsna,
  title={The RSNA-ASNR-MICCAI BraTS 2021 Benchmark on Brain Tumor Segmentation and Radiogenomic Classification},
  author={Baid, Ujjwal and Ghodasara, Satyam and Bilello, Michel and Mohan, Suyash and Calabrese, Evan and Colak, Errol and Farahani, Keyvan and Kalpathy-Cramer, Jayashree and Kitamura, Felipe C and Pati, Sarthak and others},
  journal={arXiv preprint arXiv:2107.02314},
  year={2021}
}

@article{menze2014multimodal,
  title={The multimodal brain tumor image segmentation benchmark (BRATS)},
  author={Menze, Bjoern H and Jakab, Andras and Bauer, Stefan and Kalpathy-Cramer, Jayashree and Farahani, Keyvan and Kirby, Justin and Burren, Yuliya and Porz, Nicole and Slotboom, Johannes and Wiest, Roland and others},
  journal={IEEE transactions on medical imaging},
  volume={34},
  number={10},
  pages={1993--2024},
  year={2014},
  publisher={IEEE}
}

@article{bakas2017advancing,
  title={Advancing the cancer genome atlas glioma MRI collections with expert segmentation labels and radiomic features},
  author={Bakas, Spyridon and Akbari, Hamed and Sotiras, Aristeidis and Bilello, Michel and Rozycki, Martin and Kirby, Justin S and Freymann, John B and Farahani, Keyvan and Davatzikos, Christos},
  journal={Scientific data},
  volume={4},
  number={1},
  pages={1--13},
  year={2017},
  publisher={Nature Publishing Group}
}

@article{bakas2017segmentation1,
  title={Segmentation labels and radiomic features for the pre-operative scans of the TCGA-LGG collection},
  author={Bakas, Spyridon and Akbari, Hamed and Sotiras, Aristeidis and Bilello, Michel and Rozycki, Martin and Kirby, Justin and Freymann, John and Farahani, Keyvan and Davatzikos, Christos},
  journal={The cancer imaging archive},
  volume={286},
  year={2017}
}

@article{bakas2017segmentation,
  title={Segmentation labels and radiomic features for the pre-operative scans of the TCGA-GBM collection. The Cancer Imaging Archive},
  author={Bakas, Spyridon and Akbari, Hamed and Sotiras, Aristeidis and Bilello, Michel and Rozycki, Martin and Kirby, Justin and Freymann, John and Farahani, Keyvan and Davatzikos, Christos},
  journal={Nat Sci Data},
  volume={4},
  pages={170117},
  year={2017}
}

@article{heller2023kits21,
      title={The KiTS21 Challenge: Automatic segmentation of kidneys, renal tumors, and renal cysts in corticomedullary-phase CT}, 
      author={Nicholas Heller and Fabian Isensee and Dasha Trofimova and Resha Tejpaul and Zhongchen Zhao and Huai Chen and Lisheng Wang and Alex Golts and Daniel Khapun and Daniel Shats and Yoel Shoshan and Flora Gilboa-Solomon and Yasmeen George and Xi Yang and Jianpeng Zhang and Jing Zhang and Yong Xia and Mengran Wu and Zhiyang Liu and Ed Walczak and Sean McSweeney and Ranveer Vasdev and Chris Hornung and Rafat Solaiman and Jamee Schoephoerster and Bailey Abernathy and David Wu and Safa Abdulkadir and Ben Byun and Justice Spriggs and Griffin Struyk and Alexandra Austin and Ben Simpson and Michael Hagstrom and Sierra Virnig and John French and Nitin Venkatesh and Sarah Chan and Keenan Moore and Anna Jacobsen and Susan Austin and Mark Austin and Subodh Regmi and Nikolaos Papanikolopoulos and Christopher Weight},
      year={2023},
      eprint={2307.01984},
      archivePrefix={arXiv},
      primaryClass={cs.CV}
}

@article{BILIC2023102680,
title = {The Liver Tumor Segmentation Benchmark (LiTS)},
journal = {Medical Image Analysis},
volume = {84},
pages = {102680},
year = {2023},
issn = {1361-8415},
doi = {https://doi.org/10.1016/j.media.2022.102680},
url = {https://www.sciencedirect.com/science/article/pii/S1361841522003085},
author = {Patrick Bilic and Patrick Christ and Hongwei Bran Li and Eugene Vorontsov and Avi Ben-Cohen and Georgios Kaissis and Adi Szeskin and Colin Jacobs and Gabriel Efrain Humpire Mamani and Gabriel Chartrand and Fabian Lohöfer and Julian Walter Holch and Wieland Sommer and Felix Hofmann and Alexandre Hostettler and Naama Lev-Cohain and Michal Drozdzal and Michal Marianne Amitai and Refael Vivanti and Jacob Sosna and Ivan Ezhov and Anjany Sekuboyina and Fernando Navarro and Florian Kofler and Johannes C. Paetzold and Suprosanna Shit and Xiaobin Hu and Jana Lipková and Markus Rempfler and Marie Piraud and Jan Kirschke and Benedikt Wiestler and Zhiheng Zhang and Christian Hülsemeyer and Marcel Beetz and Florian Ettlinger and Michela Antonelli and Woong Bae and Míriam Bellver and Lei Bi and Hao Chen and Grzegorz Chlebus and Erik B. Dam and Qi Dou and Chi-Wing Fu and Bogdan Georgescu and Xavier Giró-i-Nieto and Felix Gruen and Xu Han and Pheng-Ann Heng and Jürgen Hesser and Jan Hendrik Moltz and Christian Igel and Fabian Isensee and Paul Jäger and Fucang Jia and Krishna Chaitanya Kaluva and Mahendra Khened and Ildoo Kim and Jae-Hun Kim and Sungwoong Kim and Simon Kohl and Tomasz Konopczynski and Avinash Kori and Ganapathy Krishnamurthi and Fan Li and Hongchao Li and Junbo Li and Xiaomeng Li and John Lowengrub and Jun Ma and Klaus Maier-Hein and Kevis-Kokitsi Maninis and Hans Meine and Dorit Merhof and Akshay Pai and Mathias Perslev and Jens Petersen and Jordi Pont-Tuset and Jin Qi and Xiaojuan Qi and Oliver Rippel and Karsten Roth and Ignacio Sarasua and Andrea Schenk and Zengming Shen and Jordi Torres and Christian Wachinger and Chunliang Wang and Leon Weninger and Jianrong Wu and Daguang Xu and Xiaoping Yang and Simon Chun-Ho Yu and Yading Yuan and Miao Yue and Liping Zhang and Jorge Cardoso and Spyridon Bakas and Rickmer Braren and Volker Heinemann and Christopher Pal and An Tang and Samuel Kadoury and Luc Soler and Bram {van Ginneken} and Hayit Greenspan and Leo Joskowicz and Bjoern Menze}}

@inproceedings{nnunet,
  title={nnU-Net for Brain Tumor Segmentation},
  author={Isensee, Fabian and Maier-Hein, Klaus H},
  booktitle={Brainlesion: Glioma, Multiple Sclerosis, Stroke and Traumatic Brain Injuries: 6th International Workshop, BrainLes 2020, Held in Conjunction with MICCAI 2020, Lima, Peru, October 4, 2020, Revised Selected Papers, Part II},
  volume={12658},
  pages={118},
  year={2021},
  organization={Springer Nature}
}

@inproceedings{liu2021swin,
  title={Swin transformer: Hierarchical vision transformer using shifted windows},
  author={Liu, Ze and Lin, Yutong and Cao, Yue and Hu, Han and Wei, Yixuan and Zhang, Zheng and Lin, Stephen and Guo, Baining},
  booktitle={Proceedings of the IEEE/CVF international conference on computer vision},
  pages={10012--10022},
  year={2021}
}

@article{SAM,
  title={Segment anything},
  author={Kirillov, Alexander and Mintun, Eric and Ravi, Nikhila and Mao, Hanzi and Rolland, Chloe and Gustafson, Laura and Xiao, Tete and Whitehead, Spencer and Berg, Alexander C and Lo, Wan-Yen and others},
  journal={arXiv preprint arXiv:2304.02643},
  year={2023}
}

@article{MedSAM,
  title={Segment anything in medical images},
  author={Ma, Jun and He, Yuting and Li, Feifei and Han, Lin and You, Chenyu and Wang, Bo},
  journal={Nature Communications},
  volume={15},
  number={1},
  pages={654},
  year={2024},
  publisher={Nature Publishing Group UK London}
}

@article{SAMMED2D,
  title={Sam-med2d},
  author={Cheng, Junlong and Ye, Jin and Deng, Zhongying and Chen, Jianpin and Li, Tianbin and Wang, Haoyu and Su, Yanzhou and Huang, Ziyan and Chen, Jilong and Jiang, Lei and others},
  journal={arXiv preprint arXiv:2308.16184},
  year={2023}
}

@inproceedings{SAMMED3D,
  title={Sam-med3d: towards general-purpose segmentation models for volumetric medical images},
  author={Wang, Haoyu and Guo, Sizheng and Ye, Jin and Deng, Zhongying and Cheng, Junlong and Li, Tianbin and Chen, Jianpin and Su, Yanzhou and Huang, Ziyan and Shen, Yiqing and others},
  booktitle={European Conference on Computer Vision},
  pages={51--67},
  year={2025},
  organization={Springer}
}

@article{SAM3D,
  title={Sam3d: Segment anything model in volumetric medical images},
  author={Bui, Nhat-Tan and Hoang, Dinh-Hieu and Tran, Minh-Triet and Le, Ngan},
  journal={arXiv preprint arXiv:2309.03493},
  year={2023}
}

@article{sam_survey,
  title={Segment anything model for medical image segmentation: Current applications and future directions},
  author={Zhang, Yichi and Shen, Zhenrong and Jiao, Rushi},
  journal={Computers in Biology and Medicine},
  pages={108238},
  year={2024},
  publisher={Elsevier}
}

@inproceedings{he2022masked,
  title={Masked autoencoders are scalable vision learners},
  author={He, Kaiming and Chen, Xinlei and Xie, Saining and Li, Yanghao and Doll{\'a}r, Piotr and Girshick, Ross},
  booktitle={Proceedings of the IEEE/CVF conference on computer vision and pattern recognition},
  pages={16000--16009},
  year={2022}
}

@article{peiris2023uncertainty,
  title={Uncertainty-guided dual-views for semi-supervised volumetric medical image segmentation},
  author={Peiris, Himashi and Hayat, Munawar and Chen, Zhaolin and Egan, Gary and Harandi, Mehrtash},
  journal={Nature Machine Intelligence},
  pages={1--15},
  year={2023},
  publisher={Nature Publishing Group UK London}
}

@article{sinclair2024perivascular,
  title={Perivascular space Identification Nnunet for Generalised Usage (PINGU)},
  author={Sinclair, Benjamin and Vivash, Lucy and Moses, Jasmine and Lynch, Miranda and Pham, William and Dorfmann, Karina and Marotta, Cassandra and Koh, Shaun and Bunyamin, Jacob and Rowsthorn, Ella and others},
  journal={arXiv preprint arXiv:2405.08337},
  year={2024}
}

@article{shen2024fastsam3d,
  title={FastSAM3D: An Efficient Segment Anything Model for 3D Volumetric Medical Images},
  author={Shen, Yiqing and Li, Jingxing and Shao, Xinyuan and Romillo, Blanca Inigo and Jindal, Ankush and Dreizin, David and Unberath, Mathias},
  journal={arXiv preprint arXiv:2403.09827},
  year={2024}
}

@article{ali2024evaluating,
  title={Evaluating segment anything model (SAM) on MRI scans of brain tumors},
  author={Ali, Luqman and Alnajjar, Fady and Swavaf, Muhammad and Elharrouss, Omar and Abd-Alrazaq, Alaa and Damseh, Rafat},
  journal={Scientific reports},
  volume={14},
  number={1},
  pages={21659},
  year={2024},
  publisher={Nature Publishing Group UK London}
}

@article{msd,
  title={The medical segmentation decathlon},
  author={Antonelli, Michela and Reinke, Annika and Bakas, Spyridon and Farahani, Keyvan and Kopp-Schneider, Annette and Landman, Bennett A and Litjens, Geert and Menze, Bjoern and Ronneberger, Olaf and Summers, Ronald M and others},
  journal={Nature communications},
  volume={13},
  number={1},
  pages={1--13},
  year={2022},
  publisher={Nature Publishing Group}
}

@article{gatidis2022whole,
  title={A whole-body FDG-PET/CT dataset with manually annotated tumor lesions},
  author={Gatidis, Sergios and Hepp, Tobias and Fr{\"u}h, Marcel and La Foug{\`e}re, Christian and Nikolaou, Konstantin and Pfannenberg, Christina and Sch{\"o}lkopf, Bernhard and K{\"u}stner, Thomas and Cyran, Clemens and Rubin, Daniel},
  journal={Scientific Data},
  volume={9},
  number={1},
  pages={601},
  year={2022},
  publisher={Nature Publishing Group UK London}
}

@article{gatidis2023autopet,
  title={The autopet challenge: towards fully automated lesion segmentation in oncologic PET/CT imaging},
  author={Gatidis, Sergios and Fr{\"u}h, Marcel and Fabritius, Matthias and Gu, Sijing and Nikolaou, Konstantin and La Foug{\`e}re, Christian and Ye, Jin and He, Junjun and Peng, Yige and Bi, Lei and others},
  year={2023}
}

@dataset{wahid_2024_11199559,
  author       = {Wahid, Kareem and
                  Dede, Cem and
                  Naser, Mohamed and
                  Fuller, Clifton},
  title        = {Training Dataset for HNTSMRG 2024 Challenge},
  month        = may,
  year         = 2024,
  publisher    = {Zenodo},
  doi          = {10.5281/zenodo.11199559},
  url          = {https://doi.org/10.5281/zenodo.11199559},
}

@incollection{wahid2024overview,
  title={Overview of the Head and Neck Tumor Segmentation for Magnetic Resonance Guided Applications (HNTS-MRG) 2024 Challenge},
  author={Wahid, Kareem A and Dede, Cem and El-Habashy, Dina M and Kamel, Serageldin and Rooney, Michael K and Khamis, Yomna and Abdelaal, Moamen RA and Ahmed, Sara and Corrigan, Kelsey L and Chang, Enoch and others},
  booktitle={Challenge on Head and Neck Tumor Segmentation for MRI-Guided Applications},
  pages={1--35},
  year={2024},
  publisher={Springer}
}

@article{luo2025tumor,
  title={Tumor detection, segmentation and classification challenge on automated 3d breast ultrasound: The tdsc-abus challenge},
  author={Luo, Gongning and Xu, Mingwang and Chen, Hongyu and Liang, Xinjie and Tao, Xing and Ni, Dong and Jeong, Hyunsu and Kim, Chulhong and Stock, Raphael and Baumgartner, Michael and others},
  journal={arXiv preprint arXiv:2501.15588},
  year={2025}
}

@article{heller2019kits19,
  title={The kits19 challenge data: 300 kidney tumor cases with clinical context, ct semantic segmentations, and surgical outcomes},
  author={Heller, Nicholas and Sathianathen, Niranjan and Kalapara, Arveen and Walczak, Edward and Moore, Keenan and Kaluzniak, Heather and Rosenberg, Joel and Blake, Paul and Rengel, Zachary and Oestreich, Makinna and others},
  journal={arXiv preprint arXiv:1904.00445},
  year={2019}
}

@incollection{andrearczyk2022overview,
  title={Overview of the HECKTOR challenge at MICCAI 2022: automatic head and neck tumor segmentation and outcome prediction in PET/CT},
  author={Andrearczyk, Vincent and Oreiller, Valentin and Abobakr, Moamen and Akhavanallaf, Azadeh and Balermpas, Panagiotis and Boughdad, Sarah and Capriotti, Leo and Castelli, Joel and Cheze Le Rest, Catherine and Decazes, Pierre and others},
  booktitle={3D Head and Neck Tumor Segmentation in PET/CT Challenge},
  pages={1--30},
  year={2022},
  publisher={Springer}
}

@misc{ingrisch_2024_10990932,
  author       = {Ingrisch, Michael and
                  Dexl, Jakob and
                  Jeblick, Katharina and
                  Cyran, Clemens and
                  Gatidis, Sergios and
                  Kuestner, Thomas},
  title        = {Automated Lesion Segmentation in Whole-Body PET/CT
                   - Multitracer Multicenter generalization
                  },
  month        = apr,
  year         = 2024,
  publisher    = {Zenodo},
  doi          = {10.5281/zenodo.10990932},
  url          = {https://doi.org/10.5281/zenodo.10990932},
}

@misc{guanyu_yang_2022_6361938,
  author       = {Guanyu Yang and
                  Yuting He and
                  Pengfei Shao and
                  Yi Xu and
                  Xiaomei Zhu and
                  Jindi Kong and
                  Lizhan Xu and
                  Ziyue Jiang},
  title        = {Kidney Parsing Challenge 2022: Multi-Structure
                   Segmentation for Renal Cancer Treatment
                  },
  month        = mar,
  year         = 2022,
  publisher    = {Zenodo},
  doi          = {10.5281/zenodo.6361938},
  url          = {https://doi.org/10.5281/zenodo.6361938},
}

@article{adams2022prostate158,
  title={Prostate158-An expert-annotated 3T MRI dataset and algorithm for prostate cancer detection},
  author={Adams, Lisa C and Makowski, Marcus R and Engel, G{\"u}nther and Rattunde, Maximilian and Busch, Felix and Asbach, Patrick and Niehues, Stefan M and Vinayahalingam, Shankeeth and van Ginneken, Bram and Litjens, Geert and others},
  journal={Computers in Biology and Medicine},
  volume={148},
  pages={105817},
  year={2022},
  publisher={Elsevier}
}

@article{dorent2023crossmoda,
  title={CrossMoDA 2021 challenge: Benchmark of cross-modality domain adaptation techniques for vestibular schwannoma and cochlea segmentation},
  author={Dorent, Reuben and Kujawa, Aaron and Ivory, Marina and Bakas, Spyridon and Rieke, Nicola and Joutard, Samuel and Glocker, Ben and Cardoso, Jorge and Modat, Marc and Batmanghelich, Kayhan and others},
  journal={Medical Image Analysis},
  volume={83},
  pages={102628},
  year={2023},
  publisher={Elsevier}
}

@article{wijethilake2025crossmoda,
  title={crossMoDA Challenge: Evolution of Cross-Modality Domain Adaptation Techniques for Vestibular Schwannoma and Cochlea Segmentation from 2021 to 2023},
  author={Wijethilake, Navodini and Dorent, Reuben and Ivory, Marina and Kujawa, Aaron and Cornelissen, Stefan and Langenhuizen, Patrick and Okasha, Mohamed and Oviedova, Anna and Dong, Hexin and Kang, Bogyeong and others},
  journal={arXiv preprint arXiv:2506.12006},
  year={2025}
}

@article{zhang2025generalist,
  title={A generalist foundation model and database for open-world medical image segmentation},
  author={Zhang, Siqi and Zhang, Qizhe and Zhang, Shanghang and Liu, Xiaohong and Yue, Jingkun and Lu, Ming and Xu, Huihuan and Yao, Jiaxin and Wei, Xiaobao and Cao, Jiajun and others},
  journal={Nature Biomedical Engineering},
  pages={1--16},
  year={2025},
  publisher={Nature Publishing Group UK London}
}

@article{RADFM,
  title={Towards generalist foundation model for radiology by leveraging web-scale 2d\&3d medical data},
  author={Wu, Chaoyi and Zhang, Xiaoman and Zhang, Ya and Hui, Hui and Wang, Yanfeng and Xie, Weidi},
  journal={Nature Communications},
  volume={16},
  number={1},
  pages={7866},
  year={2025},
  publisher={Nature Publishing Group UK London}
}

@article{METASIM,
  title={Foundation model for efficient biological discovery in single-molecule time traces},
  author={Li, Jieming and Zhang, Leyou and Johnson-Buck, Alexander and Walter, Nils G},
  journal={Nature Methods},
  pages={1--12},
  year={2025},
  publisher={Nature Publishing Group US New York}
}

@article{MRIPTPCA,
  title={An MRI--pathology foundation model for noninvasive diagnosis and grading of prostate cancer},
  author={Shao, Lizhi and Liang, Chao and Yan, Ye and Zhu, Haibin and Jiang, Xiaoming and Bao, Meiling and Zang, Pan and Huang, Xiazi and Zhou, Hongyu and Nie, Pei and others},
  journal={Nature Cancer},
  pages={1--17},
  year={2025},
  publisher={Nature Publishing Group US New York}
}

@article{GPFM,
  title={A generalizable pathology foundation model using a unified knowledge distillation pretraining framework},
  author={Ma, Jiabo and Guo, Zhengrui and Zhou, Fengtao and Wang, Yihui and Xu, Yingxue and Li, Jinbang and Yan, Fang and Cai, Yu and Zhu, Zhengjie and Jin, Cheng and others},
  journal={Nature Biomedical Engineering},
  pages={1--20},
  year={2025},
  publisher={Nature Publishing Group UK London}
}

@article{ma2023towards,
  title={Towards foundation models of biological image segmentation},
  author={Ma, Jun and Wang, Bo},
  journal={Nature Methods},
  volume={20},
  number={7},
  pages={953--955},
  year={2023},
  publisher={Nature Publishing Group US New York}
}

@inproceedings{ye2023uniseg,
  title={Uniseg: A prompt-driven universal segmentation model as well as a strong representation learner},
  author={Ye, Yiwen and Xie, Yutong and Zhang, Jianpeng and Chen, Ziyang and Xia, Yong},
  booktitle={International Conference on Medical Image Computing and Computer-Assisted Intervention},
  pages={508--518},
  year={2023},
  organization={Springer}
}

@article{ravi2024sam,
  title={Sam 2: Segment anything in images and videos},
  author={Ravi, Nikhila and Gabeur, Valentin and Hu, Yuan-Ting and Hu, Ronghang and Ryali, Chaitanya and Ma, Tengyu and Khedr, Haitham and R{\"a}dle, Roman and Rolland, Chloe and Gustafson, Laura and others},
  journal={arXiv preprint arXiv:2408.00714},
  year={2024}
}

@article{ma2024segment,
  title={Segment anything in medical images and videos: Benchmark and deployment},
  author={Ma, Jun and Kim, Sumin and Li, Feifei and Baharoon, Mohammed and Asakereh, Reza and Lyu, Hongwei and Wang, Bo},
  journal={arXiv preprint arXiv:2408.03322},
  year={2024}
}

@article{zhu2024medical,
  title={Medical sam 2: Segment medical images as video via segment anything model 2},
  author={Zhu, Jiayuan and Hamdi, Abdullah and Qi, Yunli and Jin, Yueming and Wu, Junde},
  journal={arXiv preprint arXiv:2408.00874},
  year={2024}
}

@article{engelmann2025training,
  title={Training a high-performance retinal foundation model with half-the-data and 400 times less compute},
  author={Engelmann, Justin and Bernabeu, Miguel O},
  journal={Nature Communications},
  volume={16},
  number={1},
  pages={6862},
  year={2025},
  publisher={Nature Publishing Group UK London}
}

@article{campanella2025real,
  title={Real-world deployment of a fine-tuned pathology foundation model for lung cancer biomarker detection},
  author={Campanella, Gabriele and Kumar, Neeraj and Nanda, Swaraj and Singi, Siddharth and Fluder, Eugene and Kwan, Ricky and Muehlstedt, Silke and Pfarr, Nicole and Sch{\"u}ffler, Peter J and H{\"a}ggstr{\"o}m, Ida and others},
  journal={Nature Medicine},
  pages={1--9},
  year={2025},
  publisher={Nature Publishing Group US New York}
}

@inproceedings{pieper20043d,
  title={3D Slicer},
  author={Pieper, Steve and Halle, Michael and Kikinis, Ron},
  booktitle={2004 2nd IEEE international symposium on biomedical imaging: nano to macro (IEEE Cat No. 04EX821)},
  pages={632--635},
  year={2004},
  organization={IEEE}
}

@inproceedings{shen2024fastsam,
  title={Fastsam-3dslicer: A 3d-slicer extension for 3d volumetric segment anything model with uncertainty quantification},
  author={Shen, Yiqing and Shao, Xinyuan and Romillo, Blanca Inigo and Dreizin, David and Unberath, Mathias},
  booktitle={International Workshop on Foundation Models for General Medical AI},
  pages={1--9},
  year={2024},
  organization={Springer}
}

@inproceedings{zhang2023segment,
  title={Segment anything model (sam) for medical image segmentation: A preliminary review},
  author={Zhang, Leying and Deng, Xiaokang and Lu, Yu},
  booktitle={2023 IEEE international conference on bioinformatics and biomedicine (BIBM)},
  pages={4187--4194},
  year={2023},
  organization={IEEE}
}

@article{lu2025general,
  title={General lightweight framework for vision foundation model supporting multi-task and multi-center medical image analysis},
  author={Lu, Senliang and Chen, Yehang and Chen, Yuan and Li, Peijun and Sun, Junqi and Zheng, Changye and Zou, Yujian and Liang, Bo and Li, Mingwei and Jin, Qinggeng and others},
  journal={Nature Communications},
  volume={16},
  number={1},
  pages={2097},
  year={2025},
  publisher={Nature Publishing Group UK London}
}

@article{cardoso2022monai,
  title={Monai: An open-source framework for deep learning in healthcare},
  author={Cardoso, M Jorge and Li, Wenqi and Brown, Richard and Ma, Nic and Kerfoot, Eric and Wang, Yiheng and Murrey, Benjamin and Myronenko, Andriy and Zhao, Can and Yang, Dong and others},
  journal={arXiv preprint arXiv:2211.02701},
  year={2022}
}

\newpage
\clearpage
\thispagestyle{empty}
\renewcommand{\figurename}{Extended Data Fig.}
\setcounter{figure}{0}
\setcounter{table}{0}

\begin{figure*}[h!]
\scriptsize
\centering
\includegraphics[width=1.0\linewidth]{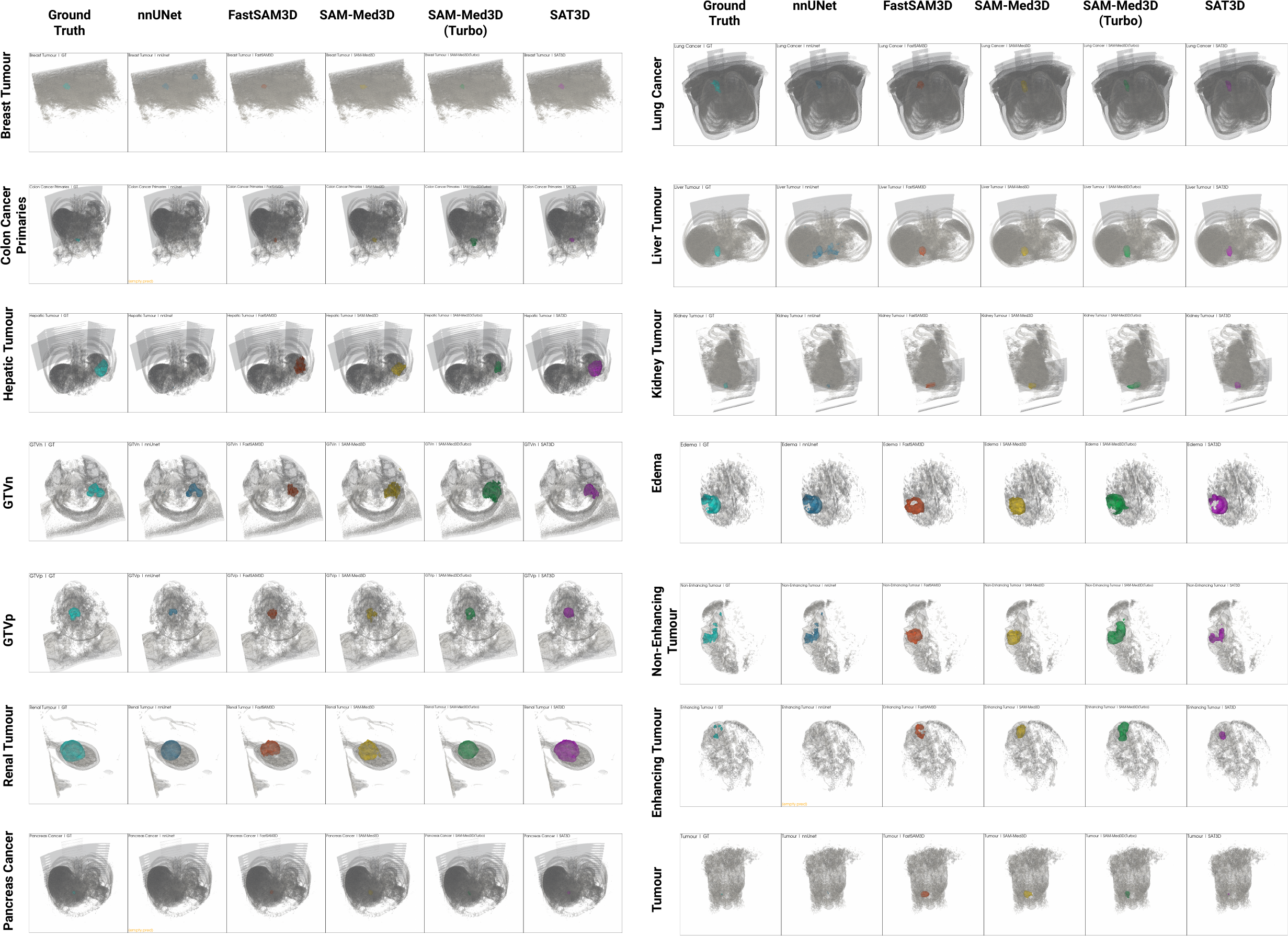} 
\caption{\textbf{Qualitative comparison of Volumetric Segmentations.} }
\label{exfig:figure_1}
\end{figure*}

\thispagestyle{empty}
\renewcommand{\figurename}{Extended Data Fig.}
\setcounter{figure}{1}

\begin{figure*}[h!]
\scriptsize
\centering
\includegraphics[width=0.95\linewidth]{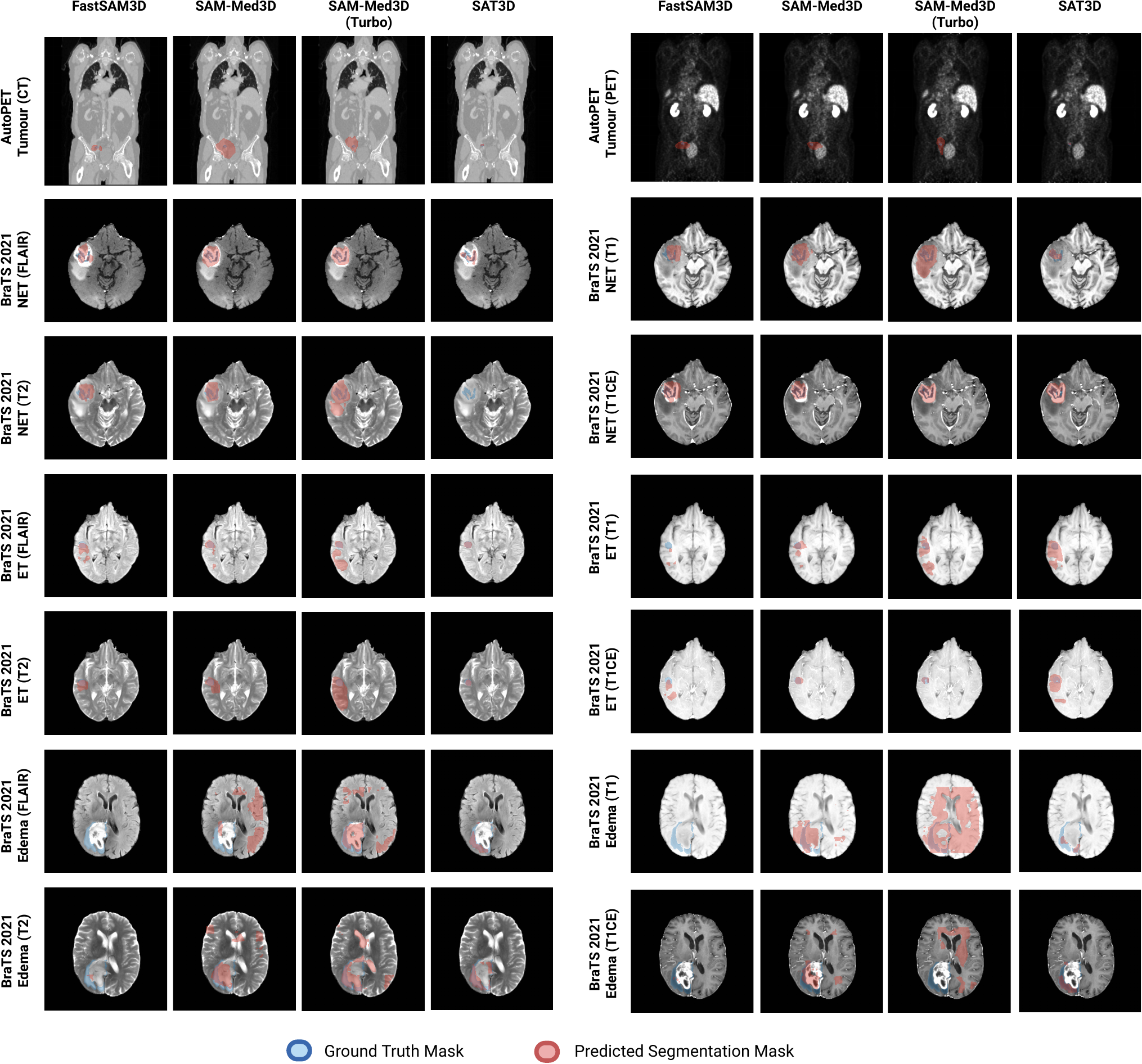} 
\caption{\textbf{Qualitative performance comparison on multi-modal imaging data.}  }
\label{exfig:figure_2}
\end{figure*}

\thispagestyle{empty}
\renewcommand{\figurename}{Extended Data Fig.}
\setcounter{figure}{2}

\begin{figure*}[h!]
\scriptsize
\centering
\includegraphics[width=0.65\linewidth]{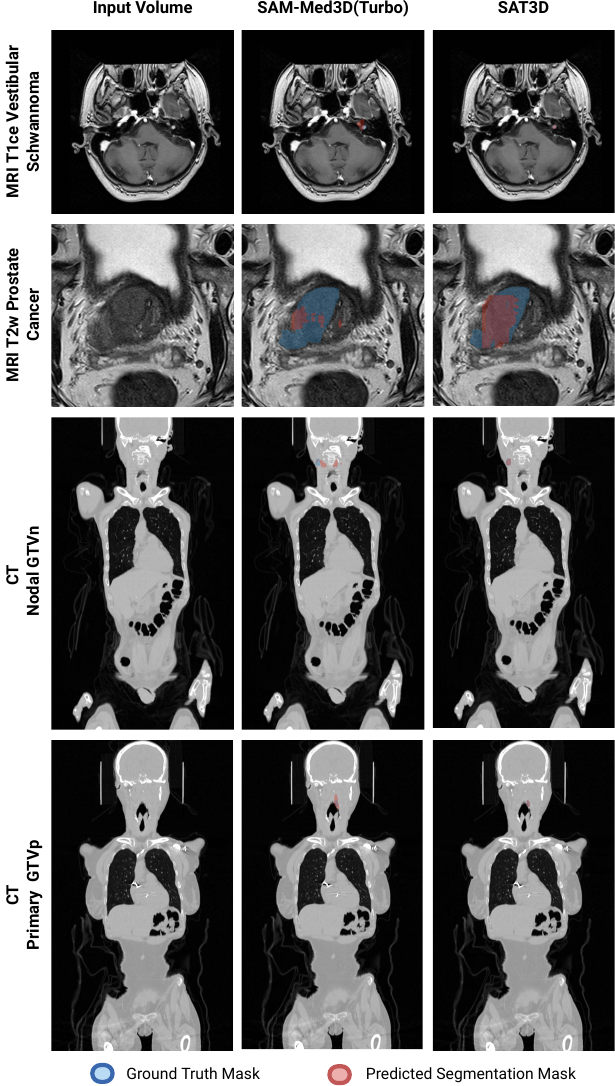} 
\caption{\textbf{Qualitative comparison of SAT3D’s segmentation performance on out-of-distribution (OOD) data.} The figure illustrates representative cases from distinct tumour types, including vestibular schwannoma, prostate cancer, and head-and-neck gross tumour volumes (GTVp and GTVn). SAT3D demonstrates robust generalisation across unseen anatomical regions and modalities, accurately delineating tumour boundaries and maintaining structural consistency in regions not encountered during training. }
\label{exfig:figure_3}
\end{figure*}

\thispagestyle{empty}
\renewcommand{\figurename}{Extended Data Fig.}
\setcounter{figure}{3}

\begin{figure*}[h!]
\scriptsize
\centering
\includegraphics[width=1\linewidth]{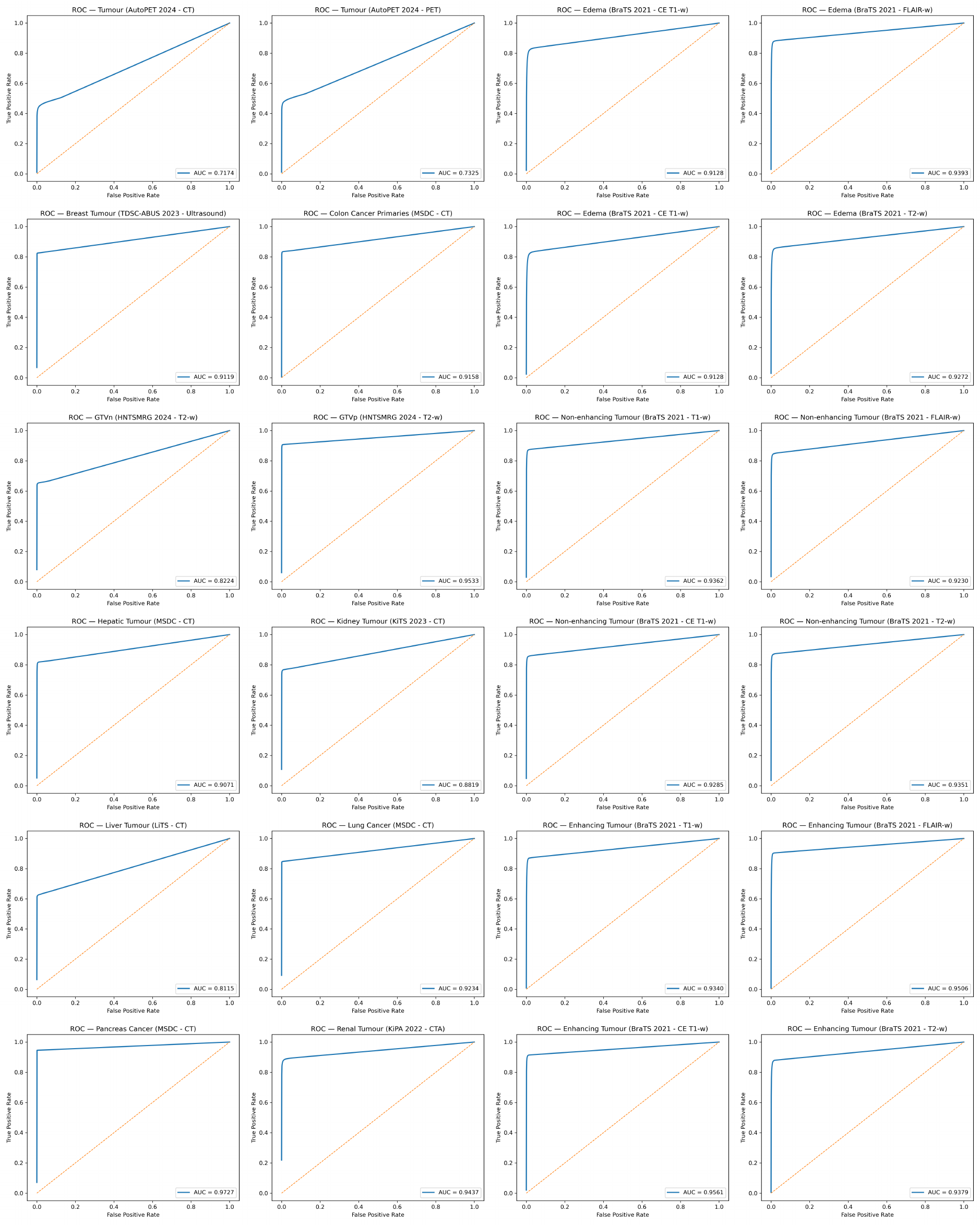} 
\caption{\textbf{Receiver Operating Characteristic (ROC) Curves}}
\label{exfig:figure_4}
\end{figure*}

\thispagestyle{empty}
\renewcommand{\tablename}{Extended Data Table}

\begin{table*}[t]
\centering
\scriptsize

\caption{Statistical significance (Wilcoxon signed-rank test) comparing SAT3D against other methods across tumour/cancer types using the DSC metric. }
\label{extab:table_1}
\resizebox{0.55\linewidth}{!}{
\begin{tabular}{lll}
\toprule
\textbf{Tumor Type} & \textbf{Comparison} & \textbf{p-value} \\
\midrule
Breast tumor & SAT3D vs nnUNet & 0.0020 \\
Breast Tumour & SAT3D vs SAM-Med3D & 6.10e-05 \\
Breast Tumour & SAT3D vs SAM-Med3D(Turbo) & 6.10e-05 \\
Breast Tumour & SAT3D vs FastSAM3D & 6.10e-05 \\
Colon Cancer Primaries & SAT3D vs nnUNet & 0.8124 \\
Colon Cancer Primaries & SAT3D vs SAM-Med3D & 0.0007 \\
Colon Cancer Primaries & SAT3D vs SAM-Med3D(Turbo) & 0.0005 \\
Colon Cancer Primaries & SAT3D vs FastSAM3D & 0.0002 \\
Edema & SAT3D vs nnUNet & 6.16e-22 \\
Edema & SAT3D vs SAM-Med3D & 2.14e-32 \\
Edema & SAT3D vs SAM-Med3D(Turbo) & 7.17e-18 \\
Edema & SAT3D vs FastSAM3D & 1.98e-32 \\
Enhancing Tumour & SAT3D vs nnUNet & 1.27e-20 \\
Enhancing Tumour & SAT3D vs SAM-Med3D & 1.26e-27 \\
Enhancing Tumour & SAT3D vs SAM-Med3D(Turbo) & 0.0005 \\
Enhancing Tumour & SAT3D vs FastSAM3D & 2.56e-27 \\
GTVn & SAT3D vs nnUNet & 0.0240 \\
GTVn & SAT3D vs SAM-Med3D & 0.0038 \\
GTVn & SAT3D vs SAM-Med3D(Turbo) & 0.0038 \\
GTVn & SAT3D vs FastSAM3D & 0.0021 \\
GTVp & SAT3D vs nnUNet & 0.9729 \\
GTVp & SAT3D vs SAM-Med3D & 0.0002 \\
GTVp & SAT3D vs SAM-Med3D(Turbo) & 0.1111 \\
GTVp & SAT3D vs FastSAM3D & 3.15e-05 \\
Hepatic Tumour & SAT3D vs nnUNet & 0.0004 \\
Hepatic Tumour & SAT3D vs SAM-Med3D & 7.05e-08 \\
Hepatic Tumour & SAT3D vs SAM-Med3D(Turbo) & 1.32e-08 \\
Hepatic Tumour & SAT3D vs FastSAM3D & 2.71e-08 \\
Kidney Tumour & SAT3D vs nnUNet & 0.0219 \\
Kidney Tumour & SAT3D vs SAM-Med3D & 5.23e-10 \\
Kidney Tumour & SAT3D vs SAM-Med3D(Turbo) & 0.0809 \\
Kidney Tumour & SAT3D vs FastSAM3D & 3.94e-09 \\
Liver Tumour& SAT3D vs nnUNet & 0.0602 \\
Liver Tumour& SAT3D vs SAM-Med3D & 0.0043 \\
Liver Tumour& SAT3D vs SAM-Med3D(Turbo) & 0.0043 \\
Liver Tumour& SAT3D vs FastSAM3D & 0.0009 \\
Lung Cancer & SAT3D vs nnUNet & 0.2754 \\
Lung Cancer & SAT3D vs SAM-Med3D & 0.0645 \\
Lung Cancer & SAT3D vs SAM-Med3D(Turbo) & 0.0273 \\
Lung Cancer & SAT3D vs FastSAM3D & 0.1055 \\
Non-Enhancing Tumour & SAT3D vs nnUNet & 0.0645 \\
Non-Enhancing Tumour & SAT3D vs SAM-Med3D & 2.39e-24 \\
Non-Enhancing Tumour & SAT3D vs SAM-Med3D(Turbo) & 3.72e-24 \\
Non-Enhancing Tumour & SAT3D vs FastSAM3D & 1.64e-28 \\
Pancreas Cancer & SAT3D vs nnUNet & 0.0003 \\
Pancreas Cancer & SAT3D vs SAM-Med3D & 6.44e-07 \\
Pancreas Cancer & SAT3D vs SAM-Med3D(Turbo) & 6.70e-08 \\
Pancreas Cancer & SAT3D vs FastSAM3D & 0.0436 \\
Renal Tumour & SAT3D vs nnUNet & 0.0674 \\
Renal Tumour & SAT3D vs SAM-Med3D & 0.0010 \\
Renal Tumour & SAT3D vs SAM-Med3D(Turbo) & 0.0830 \\
Renal Tumour & SAT3D vs FastSAM3D & 0.0010 \\
Tumour & SAT3D vs nnUNet & 1.39e-10 \\
Tumour & SAT3D vs SAM-Med3D & 2.62e-35 \\
Tumour & SAT3D vs SAM-Med3D(Turbo) & 1.30e-30 \\
Tumour & SAT3D vs FastSAM3D & 1.70e-35 \\
\bottomrule
\end{tabular}
}
\end{table*}

\clearpage
\thispagestyle{empty}
\section*{Supplementary Information}

\renewcommand{\figurename}{Extended Data Fig.}
\setcounter{figure}{0}
\setcounter{table}{0}

\section*{Evaluation Metrics.}
During evaluation, five evaluation methods were selected to evaluate the segmentation accuracy of the models, each targeting a distinct facet of quality: Dice similarity coefficient (DSC), Intersection over Union (IoU),  Relative Volume Error (RVE), 95th-percentile Hausdorff Distance (HD95), and Average Symmetric Surface Distance (ASSD). Let $\Omega \subset \mathbb{Z}^3$ be the voxel lattice, $P,G \subset \Omega$ the predicted and ground-truth foreground sets, and $|\cdot|$ voxel cardinality. Let $s=(s_x,s_y,s_z)\in \mathbb{R}_+^3$ denote voxel spacing and $V(S)=|S|\,s_xs_ys_z$ the physical volume in $\mathrm{mm}^3$. Overlap and volume metrics are dimensionless; boundary metrics are reported in millimetres (mm).

\paragraph{\textbf{Dice similarity coefficient (DSC).}}
DSC quantifies set overlap with balanced weighting of precision and recall: 
\begin{equation}
    \mathrm{DSC}(P,G)=\frac{2\,|P\cap G|}{\,|P|+|G|\,}
    \;=\;\frac{2\,\mathrm{TP}}{2\,\mathrm{TP}+\mathrm{FP}+\mathrm{FN}},
\end{equation}
where $\mathrm{TP}$, $\mathrm{FP}$ and $\mathrm{FN}$ are voxel counts in the contingency table. $\mathrm{DSC}\in[0,1]$, attaining $1$ if and only if $P=G$ and $0$ when $P\cap G=\varnothing$ with at least one of $P,G$ non–empty. It is symmetric in $(P,G)$, invariant to rigid translations, and particularly sensitive to mismatch on small or thin structures.

\paragraph{\textbf{Intersection over Union (IoU; Jaccard index).}}
The Jaccard index—also known in segmentation as Intersection over Union (IoU)—is a normalised set similarity defined as the size of the overlap relative to the size of the union:
\begin{equation}
    \mathrm{IoU}(P,G)=\frac{|P\cap G|}{\,|P|+|G|-|P\cap G|\,}
    \;=\;\frac{\mathrm{TP}}{\mathrm{TP}+\mathrm{FP}+\mathrm{FN}}.
\end{equation}
$\mathrm{IoU}\in[0,1]$, with the same extrema and invariances as DSC. The two are strictly monotonically related,
\[
\mathrm{DSC}=\frac{2\,\mathrm{IoU}}{1+\mathrm{IoU}},\qquad
\mathrm{IoU}=\frac{\mathrm{DSC}}{2-\mathrm{DSC}},
\]
and thus induce identical rankings while providing different numeric scales.

\paragraph{\textbf{Relative volume error (RVE).}}
RVE isolates volumetric bias—over- or under-segmentation—independent of spatial arrangement:
\begin{equation}
    \mathrm{RVE}(P,G)=\frac{\bigl|V(P)-V(G)\bigr|}{V(G)}.
\end{equation}
It satisfies $\mathrm{RVE}\in[0,\infty)$, equals $0$ if and only if $V(P)=V(G)$, and is naturally interpreted as a proportion (e.g., $0.10$ indicates a $10\%$ volume discrepancy). Being a ratio, RVE is scale-invariant to uniform rescaling of voxel spacing. As it disregards location and shape, it complements overlap and boundary metrics.

\paragraph{\textbf{95th–percentile Hausdorff distance (HD95).}}
Hausdorff distance probes near–worst–case boundary discrepancy. To temper sensitivity to isolated outliers, we report the \emph{robust} (area–weighted) 95th percentile in each direction and take the symmetric maximum. Let $\Sigma_P$ and $\Sigma_G$ denote the boundaries (in $\mathbb{R}^3$) induced by $P$ and $G$, and define directed sets
\[
D(P\!\to\!G)=\bigl\{\min_{y\in \Sigma_G}\|x-y\|_2:\; x\in \Sigma_P\bigr\},\quad
D(G\!\to\!P)=\bigl\{\min_{x\in \Sigma_P}\|y-x\|_2:\; y\in \Sigma_G\bigr\}.
\]
With surfel–area weighting to account for anisotropic spacing, we compute
\[
\mathrm{HD}_{95}(P,G)=\max\!\left\{
\operatorname{Q}_{0.95}^{(\text{area})}\!\bigl(D(P\!\to\!G)\bigr),\;
\operatorname{Q}_{0.95}^{(\text{area})}\!\bigl(D(G\!\to\!P)\bigr)
\right\}\quad[\mathrm{mm}].
\]
where $\operatorname{Q}_{0.95}^{(\text{area})}$ denotes the area‑weighted 95th percentile.
HD95 is non–negative, equals $0$ if and only if the boundaries coincide, and is most influenced by localised extreme deviations (e.g., spurious protrusions or missed satellite regions).

\paragraph{\textbf{Average symmetric surface distance (ASSD).}}
ASSD reflects the typical boundary displacement by using area–weighted means in both directions:
\[
\mathrm{ASSD}(P,G)=\tfrac{1}{2}\!\left(
\frac{\sum_{x\in \Sigma_P} a(x)\,\min_{y\in \Sigma_G}\|x-y\|_2}{\sum_{x\in \Sigma_P} a(x)}\;+\;
\frac{\sum_{y\in \Sigma_G} a(y)\,\min_{x\in \Sigma_P}\|y-x\|_2}{\sum_{y\in \Sigma_G} a(y)}
\right)\quad[\mathrm{mm}],
\]
where $a(\cdot)$ denotes the surfel (contour length in 2D; surface area in 3D) induced by voxel spacing.
It is non–negative, equals $0$ if and only if the boundaries coincide, and—relative to HD95—is less sensitive to isolated outliers but responsive to systematic offsets (e.g., uniform inward or outward shifts).

\paragraph{\textbf{Summary and interpretive guidance.}}
The five metrics jointly characterise overlap (DSC, IoU), size agreement (RVE), and geometric fidelity (HD95, ASSD). Higher values indicate better performance for DSC/IoU, whereas lower values are preferable for RVE/HD95/ASSD. Reporting both overlap and boundary distances (in mm) ensures sensitivity to anisotropic sampling and to boundary placement, providing a balanced and physically interpretable account of segmentation quality.

\paragraph{\textbf{Implementation notes.}}
Prior to evaluation, predictions were reoriented to match the ground-truth image orientation and resampled onto the ground-truth grid using nearest-neighbour interpolation. All distances are reported in millimetres based on the ground-truth voxel spacing. Boundary statistics (HD95, ASSD) follow the DeepMind \texttt{surface-distance} implementation\footnote{\url{https://github.com/google-deepmind/surface-distance}}, which applies surfel–area weighting to handle anisotropic sampling and computes the robust Hausdorff distance via per-direction 95th percentiles.

\section*{Comparison methods.}

We benchmark three prompt-conditioned vision foundation models and one fully supervised volumetric baseline. To ensure comparability, the task definition and evaluation metrics (as above) are held fixed. Methods that accept point prompts are evaluated with a pre-specified \emph{nominal prompt budget} of $K$ foreground points per case ($K\in\{5,10,15,20\}$). The supervised baseline receives no prompts. If a method yields multiple candidate masks under the same budget (e.g., different configurations of the $K$ points), we report, per case, the candidate attaining the highest Dice coefficient, which provides an upper bound for that budget. Only foreground points are used; no background points, scribbles or boxes are employed.

\begin{itemize}
    \item \textbf{SAM-Med3D~\cite{SAMMED3D}.} A prompt-conditioned, foundation-style approach that adapts the Segment Anything paradigm to three-dimensional medical imaging. It consumes sparse foreground points indicating the target region and returns a volumetric mask conditioned on these cues. The method is designed for broad transfer across anatomies and modalities and is reported at the specified prompt budgets $K$ as a representative high-capacity prompt-conditioned baseline.
    \item \textbf{SAM-Med3D (Turbo)~\cite{SAMMED3D}.} A computationally optimised variant of SAM-Med3D with reduced latency and memory footprint while preserving the same prompting interface and intended behaviour. It is assessed under the identical nominal prompt budget as SAM-Med3D so that differences in reported performance primarily reflect efficiency-oriented design choices rather than changes to the prompting regime. This variant is intended for time-sensitive use cases in which rapid updates are desirable without materially altering the segmentation semantics.
    \item \textbf{FastSAM3D~\cite{shen2024fastsam3d}.} A prompt-conditioned model prioritising responsiveness to sparse foreground points and favourable accuracy–latency trade-offs in volumetric settings. It targets efficient inference on typical clinical volumes while retaining the ability to incorporate a limited number of foreground cues. Evaluation under the nominal prompt budget emphasises how well the method balances prompt-driven adaptability with computational efficiency across diverse anatomies and imaging modalities.
    \item \textbf{nnUNet~\cite{nnunet}.} A dataset-adaptive, fully supervised baseline trained on labelled data and operating without prompts. It produces a single volumetric prediction per case, providing a strong reference for prompt-free performance. Because it does not rely on external cues at test time, nnUNet anchors the comparison by indicating what can be achieved in the label-rich regime, against which the prompt-conditioned approaches (evaluated at $K$ points) can be contextualised.

\end{itemize}

\section*{Dataset Details}
Table~\ref{tab:supp_train_datasets} below lists the official challenge landing pages and details that correspond to the datasets shown in the \textbf{Fig.} 2. For compactness, the single \emph{MSDC} row aggregates four tumour-centric tasks: Lung, Pancreas, Colon, and Hepatic Vessels, which are counted separately in the main text; together with AutoPET~2024, HNTSMRG~2024, TDSC\textendash ABUS~2023, KiPA~2022, KiTS~2023, LiTS, and BraTS~2021, these constitute the eleven in-domain training datasets. \emph{HECKTOR~2022}, \emph{Prostate158} and \emph{CrossMoDA~2022} are included here for completeness but are used solely as an out-of-distribution evaluation set. Where multiple mirrors or versions exist, we cite the most stable public landing page used in our experiments.

\begin{table*}[ht!]
\centering
\caption{Datasets Details.}
\label{tab:supp_train_datasets}
\scriptsize
\resizebox{1\linewidth}{!}{
\begin{tabular}{p{0.83\linewidth} p{0.42\linewidth}}
\toprule
\textbf{Dataset} & \textbf{Link} \\
\midrule
Automated Lesion Segmentation in Whole-Body PET/CT (AutoPET 2024) & \url{https://autopet-iii.grand-challenge.org/} \\ 
Head and Neck Tumor Segmentation for MR-Guided Applications (HNTSMRG 2024) & \url{https://hntsmrg24.grand-challenge.org/} \\ 
Tumor Detection, Segmentation and Classification Challenge on Automated 3D Breast Ultrasound (TDSC-ABUS 2023) & \url{https://tdsc-abus2023.grand-challenge.org/} \\ 
Kidney PArsing Challenge 2022 (KiPA 2022) & \url{https://kipa22.grand-challenge.org/} \\ 
Kidney Tumor Segmentation Challenge 2023 (KiTS 2023) & \url{https://kits-challenge.org/kits23/} \\ 
Liver Tumor Segmentation Challenge (LiTS) & \url{http://medicaldecathlon.com/} \\ 
Medical Segmentation Decathlon Challenge (MSDC) & \url{http://medicaldecathlon.com/} \\ 
Brain Tumour Segmentation Challenge (BraTS 2021) & \url{https://www.synapse.org/Synapse:syn51156910/wiki/622351} \\ 
HEad and neCK TumOR segmentation and outcome prediction in PET/CT images (HECKTOR 2022) & \url{https://hecktor.grand-challenge.org/} \\ 
Prostate158 & \url{https://github.com/kbressem/prostate158} \\ 
Cross-Modality Domain Adaptation Challenge (CrossMoDA 2022) & \url{https://crossmoda2022.grand-challenge.org/} \\
\bottomrule 
\end{tabular}
}
\end{table*}

\paragraph{Metabolically Active Tumour Lesions (AutoPET 2024):}
The AutoPET dataset is a large-scale, multi-centre resource designed to advance automated analysis of whole-body PET/CT imaging for oncologic applications~\cite{gatidis2022whole}. Unlike localised or single-organ datasets, whole-body PET/CT data pose significant challenges due to their multimodal nature (PET + CT), wide anatomical coverage, and high variability in tumour morphology. Furthermore, creating high-quality training labels for such data requires expert annotation by experienced radiologists, making manual segmentation both time-consuming and resource-intensive.
To address the lack of reproducible benchmarks and facilitate research in this domain, the AutoPET Challenge~\cite{gatidis2023autopet}, held as part of MICCAI 2022, released a public dataset comprising 1,014 annotated whole-body PET/CT scans for training. The primary task was the automated segmentation of metabolically active tumour lesions in 18F-FDG PET/CT scans. In the follow-up AutoPET-III(2024) challenge~\cite{ingrisch_2024_10990932}, the dataset was further expanded by introducing 597 PSMA-PET/CT scans, resulting in a total of 1,611 whole-body PET/CT studies.
The AutoPET dataset is publicly available via The Cancer Imaging Archive (TCIA) and is intended to support the development of robust, scalable models that can automate lesion segmentation across the full body, thereby reducing clinical burden and facilitating routine use of PET/CT biomarkers like metabolic tumour volume (MTV) and total lesion glycolysis (TLG).

\paragraph{Head \& Neck Tumour Segmentation (HNTSMRG 2024):}
The Head and Neck Tumour Segmentation for MR-Guided Applications (HNTSMRG) 2024 dataset comprises T2-weighted anatomical sequences of the head and neck region collected at The University of Texas MD Anderson Cancer Center (MDACC)~\cite{wahid_2024_11199559,wahid2024overview}. This dataset includes both fat-suppressed and non-fat-suppressed images, with all patients immobilised using a thermoplastic mask during imaging. The raw images, extracted from the Evercore institutional imaging repository, cover scans taken 1-3 weeks before the start of radiotherapy (pre-RT) and 2-4 weeks into radiotherapy (mid-RT). Each patient’s pre-RT and mid-RT image pairs consistently feature either fat-suppressed or non-fat-suppressed sequences. The dataset focuses on patients with histologically confirmed head and neck cancer (HNC), primarily oropharyngeal cancer (OPC) or cancer of unknown primary, who underwent radiotherapy at MDACC. It includes primary gross tumour volumes (GTVp) and metastatic lymph nodes (GTVn), with a variable number per patient. Multiple expert observers (3 to 4 physicians) independently segmented the GTVp and GTVn structures on both pre-RT and mid-RT scans. The dataset includes anonymised DICOM files converted to NIfTI format for user convenience, with images cropped from the top of the clavicles to the bottom of the nasal septum to ensure consistent field views and remove identifiable facial structures. Both the training and test sets reflect real-world cases and are partitioned to maintain similar distributions based on characteristics like image fat-suppression status, tumour response, and TNM staging.

\paragraph{Breast Tumour Segmentation (TDSC-ABUS 2023):}
Breast cancer is one of the leading causes of death among women globally, and early detection is crucial in reducing mortality rates. Automated 3D Breast Ultrasound is a newer approach to breast screening, offering advantages over handheld mammography, such as safety, speed, and higher detection rates of breast cancer, making it a promising method that could become prevalent worldwide in the coming years.
The TDSC-ABUS dataset comprises 200 3D volumes with refined tumour labels, obtained from an Automated 3D Breast Ultrasound (ABUS) system (Invenia ABUS, GE Healthcare) at Harbin Medical University Cancer Hospital in Harbin, China~\cite{luo2025tumor}. An experienced radiologist labelled and verified these data. The image sizes range between 843 $\times$ 546 $\times$ 270 and 865 $\times$ 682 $\times$ 354, with a pixel spacing of 0.200 mm and 0.073 mm, and a slice spacing of approximately 0.475674 mm. 
The dataset addresses three fundamental tasks in medical image analysis: tumour segmentation, classification, and detection. These tasks are particularly challenging on 3D ABUS volumes due to the large variations in tumour size and shape, irregular and ambiguous tumour boundaries, and a low signal-to-noise ratio. Moreover, the scarcity of publicly accessible ABUS datasets with well-labelled tumours has hindered the development of effective breast tumour segmentation, classification, and detection systems.

\paragraph{Kidney PArsing Challenge (KiPA 2022):}
The Kidney PArsing Challenge (KiPA22) is a MICCAI 2022 challenge dataset for multi-structure kidney parsing on contrast-enhanced CT angiography (CTA), designed to support surgery-based renal cancer care, where accurate modelling of perirenal anatomy supports surgery-based treatment planning~\cite{guanyu_yang_2022_6361938}. Each case provides voxel-wise annotations of four structures: kidney parenchyma, renal tumour, renal artery, and renal vein, reflecting the anatomical context required to localise tumour-feeding branches and to delineate venous drainage. The cohort exhibits pronounced heterogeneity, as tumours span multiple histologic subtypes and display wide variation in size and location. Arteries and veins are thin, tortuous, and of low contrast against the surrounding tissues (particularly for veins), making vessel boundaries intrinsically ambiguous. To support reproducible training while preventing label leakage, ground-truth masks are released for the 70 training cases. In contrast, the remaining cases are distributed as unlabeled images (30 open-test and 30 held-out). The dataset thus introduces CTA into our training corpus, supplying tumour annotations on angiographic images for training and evaluation.

\paragraph{Kidney Tumour Segmentation (KiTS 2023):}
The 2023 Kidney and Kidney Tumour Segmentation Challenge (KiTS23) is a competition designed to advance the development of automatic semantic segmentation systems for kidneys, renal tumours, and renal cysts~\cite{heller2023kits21}. This dataset is a refined dataset of the third iteration of the KiTS challenge, which was previously held in 2019 and 2021~\cite{heller2019kits19,heller2023kits21}. Kidney cancer is diagnosed in over 430,000 individuals annually, leading to approximately 180,000 deaths each year. Additionally, kidney tumours are identified in an even larger number of cases, and it is often not possible to determine radiographically whether a tumour is malignant or benign. Among those tumours presumed to be malignant, many are slow-growing and indolent, resulting in active surveillance becoming an increasingly popular management strategy for small renal masses. Despite this, the progression to metastatic disease remains a serious concern, underscoring the need for systems that can objectively and reliably characterise kidney tumour images to stratify risk and predict treatment outcomes. For nearly five years, KiTS has curated and expanded a publicly available, multi-institutional cohort of hundreds of segmented CT scans depicting kidney tumours, accompanied by comprehensive anonymised clinical data for each case. This dataset has served as a high-quality benchmark for 3D semantic segmentation methods and a valuable resource for translational research in kidney tumour radiomics. 
This dataset features an expanded training set with 489 cases and a test set comprising 110 cases.

\paragraph{Liver Tumour Segmentation (LiTS):}
The liver is the largest solid organ in the human body, playing a crucial role in metabolism and digestion. The image data for the LiTS challenge are collected from seven clinical sites all over the world, including (a) Rechts der Isar Hospital, the Technical University of Munich in Germany, (b) Radboud University Medical Center, the Netherlands, (c) Polytechnique Montréal and CHUM Research Center in Canada, (d) Sheba Medical Center in Israel, (e) the Hebrew University of Jerusalem in Israel, (f) Hadassah University Medical Center in Israel, and (g) IRCAD in France~\cite{BILIC2023102680}.  The LiTS benchmark dataset comprises 201 computed tomography images of the abdomen, with 194 CT scans containing lesions. All data are anonymised, and the images have been visually reviewed to exclude the presence of personal identifiers. The only processing applied to the images is transforming them into a unified NIfTI format using NiBabel in Python. Out of 201 image data, 131 are training sets, and 70 are testing sets.

\paragraph{Medical Segmentation Decathlon (MSD):}
The Medical Segmentation Decathlon (MSD) includes several datasets specifically curated for tumour segmentation across diverse anatomical regions and imaging modalities, designed to benchmark generalisation in medical image analysis~\cite{msd}.

\begin{enumerate}
\item \textbf{Lung (CT):} Includes 96 preoperative CT scans from patients with non-small cell lung cancer, focused on identifying small, sparsely located lung tumours within large anatomical volumes, with limited visual cues.

\item \textbf{Pancreas (CT):} Provides 420 portal-venous phase CT scans from patients undergoing resection of pancreatic masses. Annotations cover both pancreatic parenchyma and tumour (including cystic or solid lesions), with significant label imbalance and subtle tumour margins.

\item \textbf{Colon (CT):} Offers 190 CT scans of patients with primary colon cancer. The dataset focuses on segmenting colon cancer primaries, characterised by heterogeneous appearance, irregular morphology, and frequent occlusion.

\item \textbf{Hepatic Vessels (CT):} Comprises 443 contrast-enhanced CT scans with labels for both hepatic vessels and liver tumours. The juxtaposition of tubular vascular structures and irregular tumours in a crowded anatomical environment makes this dataset particularly demanding.
\end{enumerate}

These datasets collectively cover a wide range of tumour sites, lesion types, and imaging conditions, making MSD a unique and valuable benchmark for training and evaluating general-purpose or tumour-specialised segmentation models.

\paragraph{Brain Tumour Segmentation (BraTS) Challenge Dataset 2021:}
The BraTS (Brain Tumour Segmentation) 2021 challenge dataset is a collection of multi-institutional, multi-parametric MRI scans of brain gliomas, used to advance research in brain tumour segmentation~\cite{baid2021rsna,menze2014multimodal,bakas2017advancing,bakas2017segmentation,bakas2017segmentation1}. It is the largest and most diverse retrospective cohort of glioma patients made publicly available for the challenge. The dataset includes ground-truth annotations of tumour sub-regions, allowing for the quantitative evaluation of segmentation methods. The BraTS 2021 dataset is an update of the previous BraTS 2020 dataset, featuring more routine clinically acquired mpMRI scans from institutions not previously involved in BraTS. It includes a large number of patients, 2,040 scans in total, split into 1,251 training cases with labels, 219 validation cases without ground truth, and 570 hidden test cases used for final ranking. This dataset includes brain MRI scans of adult brain glioma patients, comprising four structural modalities (\ie, T1-weighted, T1 Contrast Enhanced, T2-weighted, FLAIR) and associated manually generated ground truth labels for each tumour sub-region (enhancement, necrosis, oedema).

\section*{Out of Distribution Dataset Details.}

\paragraph{Head \& Neck Tumour Dataset (HECKTOR 2022):}
The HECKTOR 2022 dataset, a challenge dataset for Head and Neck Tumour segmentation and outcome prediction, was organised as a satellite event of the 25th International Conference on Medical Image Computing and Computer Assisted Intervention (MICCAI) 2022~\cite{andrearczyk2022overview}. This dataset includes histologically confirmed oropharyngeal head and neck (H\&N) cancer patients who underwent radiotherapy and/or chemotherapy. The data consists of FDG-PET and low-dose non-contrast-enhanced CT images of the H\&N region, collected from nine centres using combined PET/CT scanners. The dataset consists of ground truth tumour contours provided for training cases, classified as background (0), GTVp (Class 1), and GTVn (Class 2). In this work, we use only the CT scans for the out-of-distribution evaluation. 

\paragraph{Prostate158}
Prostate158 is a curated, expert-annotated biparametric prostate MRI dataset acquired at a single German university hospital (Charit\'e University Hospital Berlin; 3.0\,T, Siemens VIDA/Skyra) between February 2016 and January 2020~\cite{adams2022prostate158}. Each study includes axial T2-weighted (T2w) imaging and diffusion-weighted imaging (DWI) with apparent diffusion coefficient (ADC) maps. Pixel-wise labels (NIfTI) are provided for the central gland (central\,+\,transition zones), the peripheral zone, and lesions suspicious for clinically significant cancer (PI\textendash RADS\,\(\geq\)4), with histopathological confirmation available for all cancerous lesions. To support robust benchmarking, the dataset is split into 139 training and 19 test cases; the test subset carries independent annotations from two board-certified radiologists to quantify inter-observer variability. All images are fully de-identified and underwent harmonising pre-processing (bias-field correction, resampling to unify orientation/direction/spacing, and field-of-view cropping) to standardise inputs across sequences. In this work, we use the Prostate158 dataset's 134 cases from T2w MRI scans, exclusively for out-of-distribution evaluation of prostate anatomy and lesion segmentation.

\paragraph{Cross-Modality Domain Adaptation for Medical Image Segmentation (CrossMoDA 2022)}
CrossMoDA~2022 is a multi-centre skull-base MRI benchmark targeting the segmentation of vestibular schwannoma (VS) and the bilateral cochleae in patients planned for stereotactic radiosurgery~\cite{dorent2023crossmoda,wijethilake2025crossmoda}. The dataset couples contrast-enhanced T1-weighted (ceT1) and high-resolution T2-weighted (hrT2) acquisitions drawn from two Gamma Knife centres (Queen Square Radiosurgery Centre, London, UK; Elisabeth‑TweeSteden Hospital, Tilburg, NL) In the 2022 release, the segmentation track (Task 1) provides voxel-wise annotations on 210 ceT1 studies for VS and cochleae; the accompanying hrT2 cohort comprises 210 unlabelled training scans, 64 validation scans, and 271 hidden-test scans used by the challenge for ranking using DSC and ASSD. In this work, we use the labelled ceT1 subset from Task 1 as an out-of-distribution evaluation set.


\section*{Reproducibility.}
In~\cref{tab:hyp_setting}, we provide supplementary material that defines all the hyperparameters required to reproduce the results presented in the main paper's Results sections. This will enable researchers to reproduce and compare our results with their own methods.

\begin{table*}[ht!]
\centering
\caption{Training Configurations \& Hyperparameters.}
\label{tab:hyp_setting}
\scriptsize
\resizebox{1\linewidth}{!}{
\begin{tabular}{ll}
\toprule
\multicolumn{2}{l}{\textbf{Data \& Transforms}}\\
\midrule
Transforms & \texttt{ToCanonical}; \texttt{CropOrPad} to $128\times128\times128$ (mask-guided); \texttt{RandomFlip} (axes: 0,1,2) \\
Normalization & \texttt{ZNormalization(masking\_method: $x{>}0$)} \\
Sampler & \texttt{DistributedSampler} (per-epoch shuffling via \texttt{set\_epoch}) \\
\midrule
\multicolumn{2}{l}{\textbf{Training}}\\
\midrule
Epochs & \texttt{500} \\
GPUs & 2 \\
Batch size & \texttt{3} (per GPU) \\
Accumulation & \texttt{20} steps \\
Num workers & \texttt{24} \\
Image Crop size & $128\times128\times128$ \\
Loss (seg) & \texttt{DiceCELoss(sigmoid=True, squared\_pred=True, reduction='mean')} \\
Adversarial & Generator loss weight $=0.01$; Critic trained with BCE (real/fake) \\
Uncertainty term & CE-based mask with critic map; threshold $T_m=0.3$; weight $=0.1$ \\
\midrule
\multicolumn{2}{l}{\textbf{Optimization}}\\
\midrule
Optimizer & \texttt{AdamW} \\
LRs & Image encoder: \texttt{lr}; Prompt encoder: \texttt{0.1*lr}; Mask decoder: \texttt{0.1*lr} \\
Base LR / WD & \texttt{lr = 8e-4}, \texttt{weight\_decay = 1e-5}; betas = (0.9, 0.999) \\
Scheduler & \texttt{CosineAnnealingLR(T\_max = 500)} (also for critic) \\
\midrule
\multicolumn{2}{l}{\textbf{Checkpointing \& Metrics}}\\
\midrule
Saves & latest / loss\_best / dice\_best (model \& critic) every epoch; step-best if Dice$>$0.9 \\
Dice computation & Threshold 0.5 on sigmoid mask; per-sample Dice averaged over batch \\
\bottomrule
\end{tabular}
}
\end{table*}



\end{document}